\documentclass{aa}
\usepackage{txfonts}
\usepackage{eepic}
\usepackage{float}
\usepackage{graphicx}
\usepackage{natbib}
\usepackage{bigstrut}
\usepackage{multirow}
\usepackage{subfigure}
\usepackage{amssymb}
\usepackage{color}
\usepackage[normalem]{ulem}
\newcommand{\WHz}{\mathrm{\;W/\sqrt{\mathrm{Hz}}}}
\newcommand{\mum}{\mathrm{\;\mu m}}

\begin{document}
 \title{A kilo-pixel imaging system for future space based far-infrared observatories using microwave kinetic inductance detectors }

\author{J.J.A. Baselmans\inst{1,2} \and J. Bueno\inst{1} \and S.J.C. Yates\inst{3} \and O. Yurduseven\inst{2} \and N. Llombart\inst{2} \and K. Karatsu\inst{2} \and A.M. Baryshev\inst{3,4} \and L. Ferrari\inst{2} \and A. Endo\inst{2,5} \and D.J. Thoen\inst{2} \and P.J. de Visser\inst{1} \and R.M.J. Janssen\inst{5,6} \and V. Murugesan\inst{1}\and E.F.C. Driessen\inst{7} \and G. Coiffard\inst{7} \and J. Martin-Pintado\inst{8} \and P. Hargrave\inst{9} \and M. Griffin\inst{9}.}
%%J.J.A. Baselmans,  J. Bueno, S.J.C. Yates, O. Yurduseven, N. Llombart, K. Karatsu, A.M. Baryshev, L. Ferrari, A. Endo, D.J. Thoen, P.J. de Visser, R.M.J. Janssen, V. Murugesan, E.F.C. Driessen, G. Coiffard, J. Martin-Pintado, P. Hargrave, M. Griffin.
\institute{
SRON - Netherlands Institute for Space Research, Sorbonnelaan 2, 3584 CA - Utrecht, The Netherlands\\  
\email{J.Baselmans@sron.nl} \and
Terahertz Sensing Group, Delft University of Technology, Mekelweg 1, 2628 CD - Delft, The Netherlands \and	%2
SRON - Netherlands Institute for Space Research, Landleven 12, 9747AD - Groningen, The Netherlands \and	 %3
Kapteyn Astronomical Institute, University of Groningen, Landleven 12, 9747 AD - Groningen, The Netherlands \and  %4
Kavli Institute of Nanoscience, Faculty of Applied Sciences, Delft University of Technology, Lorentzweg 1, 2628 CJ Delft, The Netherlands \and %5
Leiden Observatory, University of Leiden, PO Box 9513, 2300 RA Leiden, The Netherlands \and   %6
Institut de RadioAstronomie Millim\'{e}trique (IRAM), 300 Rue de la Piscine, 38406 Saint-Martin-d'H\'{e}res, Grenoble, France \and  %7
Centro de Astrobiologia, Ctra de Torrej\'{o}n a Ajalvir, km 4 28850 Torrejon de Ardoz (Madrid), Spain \and %8
Cardiff school of Physics and Astronomy, The Parade, Cardiff CF24 3AA, UK %9
}
\date{September 5 2016}

\abstract
%context
{}
%aims
{Future astrophysics and cosmic microwave background space missions operating in the far-infrared to millimetre part of the spectrum will require very large arrays of ultra-sensitive detectors in combination with high multiplexing factors and efficient low-noise and low-power readout systems. We have developed a demonstrator system suitable for such applications.}
%methods
{The system combines a 961 pixel imaging array based upon Microwave Kinetic Inductance Detectors (MKIDs) with a readout system capable of reading out all pixels simultaneously with only one readout cable pair and a single cryogenic amplifier. We evaluate, in a representative environment, the system performance in terms of sensitivity, dynamic range, optical efficiency, cosmic ray rejection, pixel-pixel crosstalk and overall yield at at an observation centre frequency of 850 GHz and 20\% fractional bandwidth.}
%results
{The overall system has an excellent sensitivity, with an average detector sensitivity $\mathrm{<NEP_{det}>=3\times10^{-19}\;\WHz}$ measured using a thermal calibration source. At a loading power per pixel of 50fW we demonstrate white, photon noise limited detector noise down to 300 mHz. The dynamic range would allow the detection of $\sim$ 1 Jy bright sources within the field of view without tuning the readout of the detectors. The expected dead time due to cosmic ray interactions, when operated in an L2 or a similar far-Earth orbit, is found to be $<$4\%. Additionally, the achieved pixel yield is 83\% and the crosstalk between the pixels is $<$-30dB.}
%conclusions
{This demonstrates that MKID technology can provide multiplexing ratios on the order of a 1000 with state-of-the-art single pixel performance, and that the technology is now mature enough to be considered for future space based observatories and experiments.}

\titlerunning{A large format imaging system for future space based far-infrared observatories}
\authorrunning{J.J.A. Baselmans et al.}
\maketitle

\section{Introduction}

About half the energy generated in the Universe since the Big Bang, from stellar radiation and accretion processes, comes to us in the far infrared (FIR) and sub-mm spectral range (0.03 - 1 mm) \citep{Dole2006}.  Access to this spectral range is therefore essential for astrophysics and cosmology as it allows us  to gain understanding of cold, distant, and dust enshrouded objects, many of which are completely invisible in other spectral ranges. Unfortunately observations are very difficult: the Earth's atmosphere is opaque over a large fraction of this spectral range, thus requiring observations from space. To reach the natural astrophysical backgrounds an observatory with an actively cooled telescope is required for a large fraction of the FIR spectral range in combination with background limited detectors. The required photon noise limited sensitivity of the detectors, NEP$_{ph}$, depends on the power absorbed per pixel in the instrument and ranges from NEP$_{ph}$$\sim$5$\times 10^{-18} \WHz$ for cosmic microwave background (CMB) instrument to NEP$_{ph}$$\sim$1$\times 10^{-20} \WHz$ for a grating spectrometer on an observatory with a 5-K telescope. For most space missions the total pixel count needed will be $\mathrm{\sim10^4}$. The combination of sensitivity and pixel count presents a major challenge for future detector systems. Recent experiments using thermal calibration sources have shown that it is possible to reach, or at least approach, the required detector sensitivities with a number of different technologies. Examples are Quantum Capacitance Detectors (QCD's)\citep{Echternach2013}, Transition Edge Sensors (TES's) \citep{Suzuki2016,Audley2016}, small-volume hot-electron bolometers \citep{Karasik2011} and Microwave Kinetic Inductance Detectors (MKIDs) \citep{Visser2014}. MKIDs, pioneered by \citet{Day2003}, are in essence superconducting resonant circuits designed to efficiently absorb radiation. They offer an attractive option to construct a large imaging system due to their intrinsic ease of using frequency division multiplexing at microwave frequencies, which allows many pixels to be read out using a single readout line. MKIDs have been operated successfully at millimetre wavelengths on the IRAM 30 m telescope \citep{NIKA2010,Nika22016} and at near infrared/optical wavelengths at Palomar \citep{Strader2013}. 
%------------------------------------------------------------------------------------------------------------------------------------------------------------------------------------
\begin{table*}
\caption{The detector requirements for the various mission concepts discussed in the text.}              % title of Table
\label{table:1}      % is used to refer this table in the text
\centering                                      % used for centering table
\begin{tabular}{l c c c c c c}          % centered columns (4 columns)
\hline\hline                        % inserts double horizontal lines
  								& $\lambda$ 	& $P_{det}$ 		& $NEP_{ph}$ 		&  $P_{det}$ for 1 Jy 	&	Time constant		&  1/f knee\\   
								& ($\mum$)	& (fW)		& $\mathrm (10^{-19}\WHz)$ 	& (fW)			&	(msec.)			& (Hz)\\    
\hline                                   % inserts single horizontal line
\textbf{Double Fourier interferometer} 	& 25 - 50 		& 0.029 			& 6.1 			&4.6				&  0.2 & 1 \\   
2 3m $\diameter$ 5 K telescopes		& 50 - 100		& 0.022 			& 3.7 			&2.3				&   &  \\   
0.5$\lambda$/D pixels 				& 100 - 200	& 0.018 			& 2.4 			&1.2				&   &   \\   
								& 200 - 400 	& 0.83 			& 10  			&0.58			&   &  	\\   
\hline
\textbf{Single dish Broadband camera} 	& 30 			& 0.053 			& 8.5 			&30				&  30 & 0.1 	 \\   
3m $\diameter$  5 K telescope			& 60			& 0.043 			& 5.2 			&15				&   & 	  \\   
0.5$\lambda$/D pixels 				& 120		& 0.030 			& 3.3 			&7.4				&   &  	 \\   
$\lambda/\Delta\lambda$=3			& 240 		& 0.041 			& 2.6 			&3.7				&   & 	 \\     
								& 400 		& 0.27 			& 5.2 			&2.2				&   &	\\    						
\hline
\textbf{Single dish Broadband camera} 	& 30 			& 0.053 			& 8.5 			&120				 &  30 & 0.1	\\   
10m $\diameter$ 25 K telescope		& 60			& 0.88 			& 24  			&60				&   &	\\   
0.5$\lambda$/D pixels 				& 120		& 24	 			& 89  			&30				&   &\\   
$\lambda/\Delta\lambda$=3			& 240 		& 77	 			& 113 			&15				&   &\\     
								& 400 		& 89	 			& 93 				&8.9				 &   & \\    							
\hline
\textbf{Single dish Grating spectrometer} 	& 30 			& 9.1$\times 10^{-5}$ 	& 0.35 			&0.052			 &  100 & 0.1 \\   
3m $\diameter$ 5 K telescope			& 60			& 7.1$\times 10^{-5}$ 	& 0.22 			&0.026			&   & \\   
0.5$\lambda$/D pixels 				& 120		& 5.2$\times 10^{-5}$ 	& 0.13 			&0.013			&   & \\   
$\lambda/\Delta\lambda$=1000			& 240 		& 6.9$\times 10^{-5}$ 	& 0.11 			&0.0065			&   &		 \\     
								& 400 		& 4.8$\times 10^{-5}$	& 0.22 			&0.039			&   &		 \\    			
\hline
\textbf{Single dish Grating spectrometer} 	& 30 			& 9.1$\times 10^{-5}$ 	& 0.35 			&0.21			&  100 & 0.1 \\   
10m $\diameter$  25 K telescope		& 60			& 1.5$\times 10^{-4}$ 	& 1.0 			&0.1			&   & \\   
0.5$\lambda$/D pixels 				& 120		& 0.044 			& 3.3 			&0.052				&   &\\   
$\lambda/\Delta\lambda$=1000			& 240 		& 0.13 			& 2.6 			&0.026				&   & \\     
								& 400 		& 0.15 			& 5.2 			&0.016				&   & \\    	
\hline
\textbf{CMB experiment} 				& 400 		& 120 			& 111			&0.57				&  5 & 0.1  \\   
2 m $\diameter$ 30 K telescope		& 600		& 110			&  85				&0.38				&   & \\   
1$\lambda$/D pixels 					& 900		& 96	 			& 65				&0.25				&   & \\   
$\lambda/\Delta\lambda$=3			& 1400 		& 107 			& 54				&0.16				&   &\\     
								& 2000 		& 123 			& 50				&0.11				&   &\\    	
								& 3000 		& 129 			& 41				&0.076				&   & \\    									
\hline                                             %inserts single line
\hline  
\end{tabular}
\tablefoot{
On top of the requirements listed in the table, all detector systems have the common requirements of: i) a cosmic ray dead time $<$20\% and ii) a pixel-pixel crosstalk (after data de-correlation) $<$-30 dB. Additionally all instruments will require on the order of several $\mathrm{10^4}$ of pixels.  }
\end{table*}
%-------------------------------------------------------------------------------------------------------------------------------------------------------------------------------------
However, up to now there has been no demonstration of a large scale detector system with sufficient sensitivity for operation in space. In this paper we report on the design, fabrication and evaluation of a kilo-pixel imaging system designed for future space-born FIR observatories that is based on a large array of MKIDs in combination with a dedicated readout system. 

The paper is organised as follows: We present in Section \ref{Sec:DetReq} a summary of the generic requirements for near- and far-future missions in the FIR and sub-mm from which we derive a set of specifications for the detector system discussed in the remainder of the text. In Section \ref{Sec:DetDes} we describe the design and fabrication of the detector array, in Section \ref{Sec:ExpSys} the experimental system and readout electronics, and in Section \ref{Sec:Exp} we describe the experimental results: We have performed a set of dedicated tests measuring i) optical efficiency, ii) sensitivity, iii) dynamic range, iv) pixel-to-pixel crosstalk, v) noise spectral dependence, vi) sensitivity to cosmic rays, and vii) pixel yield. We discuss in Section \ref{Sec:Dis} the measured results and discuss briefly the outlook for using a MKID system in a space based observatory and end with our concluding remarks in Section\ref{Sec:Con}.

\section{Detector requirements for future space missions} 
\label{Sec:DetReq}
%------------------------------------------------------------------------------------------------------------------------------------------------------------------------------------
\begin{table*}
\caption{Key requirements for the demonstrator system.}              % title of Table
\label{table:2}      % is used to refer this table in the text
\centering                                      % used for centering table
\begin{tabular}{l c c c c c c c c c c}          % centered columns (4 columns)
\hline\hline                        % inserts double horizontal lines \textbf{$\lambda/\Delta\lambda$}
 	& \textbf{MUX} 	& \boldmath$\lambda$	  & \boldmath$\lambda/\Delta\lambda$& \boldmath$\mathrm{NEP_{det}}$	&  \textbf{Absorption}		& \textbf{dynamic}	& \textbf{Cosmic Ray}	&  \textbf{Crosstalk}	& \textbf{1/f knee}& \textbf{Yield}\\  	
 				& \textbf{(factor)}	& 		 		& 							& 							&  \textbf{efficiency}			& \textbf{range}	& \textbf{dead time}		&  				& 			& \\  	 
\textbf{Baseline}	& 500			& 350 $\mum$			&  5							& 5$\times10^{-19}{\WHz}$			& $>$0.5					& $>$ 1000			& $<$30\%	 		&	$<$-20 dB		& $<$0.5Hz	&$>$60\%	\\  	
\textbf{Goal}		& 1000			& 200 $\mu$m			&  1.5						& 1$\times10^{-19}{\WHz}$			& $>$0.7					& $>$ $\mathrm{10^4}$	& $<$10\%	 		&	$<$-30 dB		& $<$0.1Hz	&$>$70\%	\\ 							
\hline                                             %inserts single line
\hline  
\end{tabular}
\end{table*}%-------------------------------------------------------------------------------------------------------------------------------------------------------------------------------------

We have examined the detector requirements for astrophysical applications through the analysis of four challenging mission concepts considered to be representative of various plausible future missions in the FIR/sub-millimetre: 
\begin{itemize}
\item A double Fourier interferometer with two cold apertures, similar to the SPIRIT mission concept \citep{Leisawitz2007};
\item A single-dish telescope actively cooled to 5K, such as the Japanese SPICA observatory; \citep{Swinyard2009} and the Far-Infrared Surveyor \citep{Bradford2015}, equipped with a wide-field camera and/or a grating spectrometer; 
\item A 10m class single-dish telescope passively cooled to 25 K such as Millimetron \citep{Millimetron2012}, equipped with a wide-field camera and/or a grating spectrometer; 
\item A fourth-generation CMB polarisation experiment such as COrE+ \citep{Core2015}. 
\end{itemize}
These concepts are intended to be indicative rather than definitive in that they generally cover the key performance parameter space. To derive the required  sensitivity of detector arrays for the various mission concepts, we modelled the instruments with realistic parameters and approximations for the telescopes, the optics and the sky background. As input for the calculations we used the models of \citet{Kelsall1998} and \citet{Benford2007} for the sky emission, which include contributions from the zodiacal light, the Cosmic Infrared Background (CIRB) and the CMB. We use the following realistic parameters to describe the total instrument efficiency:\\
The telescope emissivity is taken to be 2\% except for the CMB mission, for which 1\% is adopted. The system optical efficiency is the multiplication of several factors: For all cases we use a filter/mirror transmission efficiency of 0.45, a Lyot stop efficiency of 0.95, and a detector absorption efficiency of 0.8 together with an area fill factor of 0.8. For the interferometer, an additional beam divider efficiency of 0.5 is included, and for the CMB instrument an efficiency factor of 0.5 was added to represent the polarisation sensitivity of the detectors. We use a pixel size corresponding to instantaneous full sampling of the diffraction-limited beam (i.e. 0.5$\lambda$/D pixel side), except for the CMB experiment, for which a larger pixel size of 1.0$\lambda$/D is adopted.
%==========================================================================================================================
\begin{figure*}
\centering
\centering
\includegraphics[width=1\textwidth]{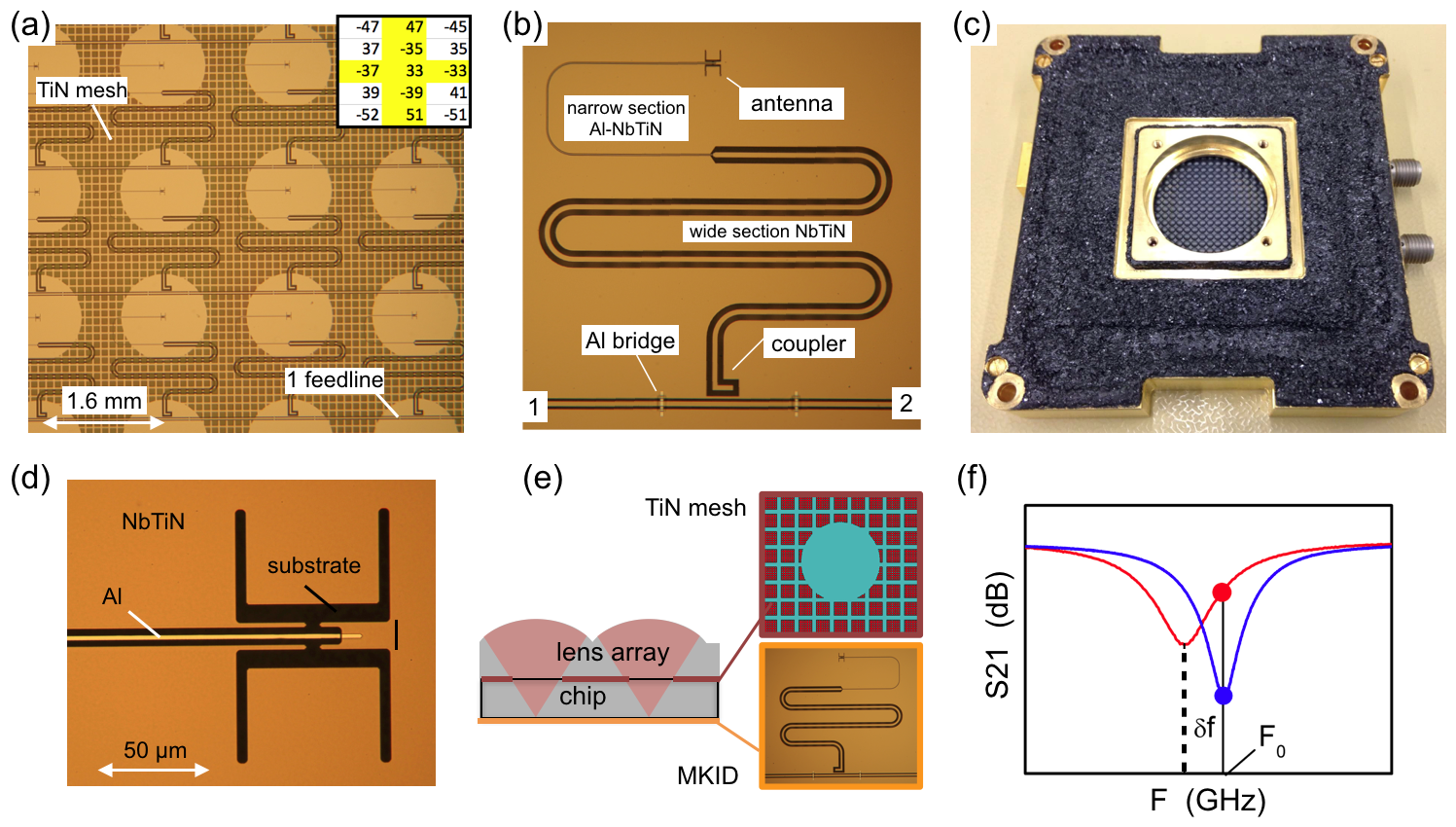}
\caption{(a) Photograph of a section of the chip, taken from the backside of the wafer where the lens array will be mounted, showing the MKIDs seen through the sapphire substrate. Also visible is the TiN mesh layer, with the holes to allow the antenna beams to couple efficiently to the lenses. Note that all meandering resonators have a slightly different length to allow them to be read out at different frequencies. (b) Zoom-in on a single MKID detector, photographed from the front side of the chip. We see the NbTiN layer and the lithographic bridges used to balance the feedline ground planes. (c) Photograph of the chip-lens array assembly in its holder, with the lens array clearly visible. In operation we place a polariser and set of bandpass filters on the circular aperture and mount the assembly inside the light-tight box of the cryostat, which is cooled to 120 mK. As a result we can only illuminate a fraction of the pixels of the array. (d) A zoom-in of panel (b) showing the antenna at the shorted end of the MKID resonator. (e) Schematic diagram of the cross section of the assembled detector array with lens array, chip and the positions of the MKIDs and the TiN mesh stray-light absorbing layer. (f) The transmission of the feedline around a single MKID measured from contact 1 to 2 in panel (b). The MKID traces a resonance dip which changes upon radiation absorption: the blue line is the equilibrium case, and the red curve corresponds to the MKID absorbing radiation. The two dots indicate the change in response of the forward scattering parameter (S21) when reading out the device with a readout tone at F$_0$.}
\label{Fig:1}
\end{figure*}
%==========================================================================================================================

We calculate, for the relevant wavelength ranges of all instrument concepts, both the power per pixel from the instrument background as well as the power from a 1-Jy source \citep{Griffin2016}: The detectors in a FIR observatory will need to be optimised for imaging very faint sources, but they should also be capable of observing much brighter sources within their instantaneous dynamic range. We also specify, based on plausible operational requirements, the maximum detector time constant and the 1/{\it f} noise requirements. The results are given in Table \ref{table:1}. The frequency range, detection bandwidth, detector time constant and 1/{\it f} knee all depend on the application. The requirements on the susceptibility to ionising radiation (cosmic rays) and pixel-pixel crosstalk are the same for all concepts. 

For the purpose of this paper we define a set of generic requirements for the detector system which we strive to achieve, which are given in in Table \ref{table:2}. The instantaneous dynamic range is derived from the ratio of a 1-Jy source power and the NEP requirements in Table \ref{table:1}. The number of pixels is driven by the requirement to be able to build systems with $\mathrm{10^4}$ pixels, for which a multiplexing factor of 500-1000 would result in a manageable amount of readout cabling and electronics; the wavelength range and NEP requirements are driven by the existing MKID technology: A NEP = 3.8$\mathrm \times10^{-19}\;\WHz$ was demonstrated by \citet{Visser2014} at 1.55 THz using amplitude readout of an aluminium MKID. A high optical efficiency was demonstrated by \citet{Janssen2013} at 350 GHz using a NbTiN-Aluminium hybrid MKID. The same publication has shown that only NbTiN-Al hybrids allow background limited performance with so-called MKID phase readout. Importantly \citet{Rantwijk2016} have demonstrated that phase readout allows the multiplexing of much more pixels than amplitude readout. This is due to the higher output signal power density from the MKIDs relative to the system noise level. Hence the hybrid design was chosen, with the challenge to combine, in a large scale system, the sensitivity from \citet{Visser2014} with the high optical efficiency and multiplexability from \citet{Janssen2013}.

\section{Detector array design and fabrication} 
\label{Sec:DetDes}
The detector array consists of a 31 x 31 (961) pixel array of MKIDs, with a pixel spacing of 1.6 mm covering an area of 49.6 x 49.6 mm on a 55 x 55 mm chip. The chip size is driven by the limiting experimental space and the pixel size and square sampling are driven by the availability of a Si lens array (see below). As stated in the previous section, we use the antenna-coupled hybrid NbTiN-Al MKID as the building block of the detector array because of its maturity and multiplexing advantage. A micrograph of a few devices of the array is shown in Fig.  \ref{Fig:1}(a), and a zoom-in on a single device is shown in Fig. \ref{Fig:1}(b). The hybrid MKID consists of a coplanar waveguide (CPW) resonator with an open and a shorted end, made out of a 500-nm thick film of NbTiN, deposited onto a 0.350-mm thick C-plane sapphire substrate using reactive magnetron sputtering in an argon-nitrogen plasma \citep{Boy2016,Thoen2016}.  We use sapphire because our fabrication yield is higher than on Si as in \citet{Janssen2013}. We read out the MKID at the first distributed resonance occurring at a frequency $F_0=\frac{c}{4L\sqrt{\epsilon_{eff}}}$. Here {\it L} is the resonator length, {\it c} the speed of light and $\epsilon_{eff}$ the effective dielectric constant of the CPW; across the array, {\it L} is changed systematically from 7.6 to 5.3 mm, resulting in F$_0$ ranging from 4.1 to 5.8 GHz. The MKID, shown in see Fig. \ref{Fig:1}(b), has a wide section over 75\% of the MKID length (linewidth = 20 $\mum$, gapwidth = 20 $\mum$) and a narrow section for 25\% of the MKID length (linewidth = gapwidth = 2 $\mum$). The wide NbTiN section near the coupler strongly reduces two-level system (TLS) noise from the device itself due to both the favourable properties of NbTiN and the large width of the CPW \citep{Gao2007, Gao2008b, Barends2009b}. The narrow section of the MKID is connected to the feed of a twin-slot antenna as shown in Fig. \ref{Fig:1}(d) and made from a NbTiN ground plane and an aluminium central line. NbTiN has a critical temperature of 15.2 K and gap frequency of 1.1 THz; for radiation at lower frequencies the surface impedance is almost purely inductive at T$<<$ Tc. On the other hand the aluminium is resistive with a sheet impedance of 0.143 $\Omega$ for frequencies exceeding 95 GHz, due to the low Tc = 1.29 K. Hence the antenna and the ground plane of the narrow CPW are lossless for 850 GHz radiation, with the result that all radiation power coupled to the antenna is absorbed in the aluminium central line. This results in the creation of quasiparticle excitations which modify the complex impedance of the aluminium at the readout frequency. This in turn changes the resonator resonant frequency and resonance shape as indicated in Fig. \ref{Fig:1}(f): The resonance feature moves to lower frequencies and becomes broader and shallower. We read out this response using a single readout tone close to F$_0$ for each resonator. The length and width of the narrow NbTiN-Al line are designed to give $>$95\% radiation absorption and negligible radiation loss within the limits of the contact lithography used in the device fabrication. Additionally, the length is minimised to reduce the device TLS noise \citep{Gao2008b}. We use aluminium for the radiation absorption due to its superior intrinsic sensitivity as demonstrated by \citet{Visser2014}.

Efficient radiation coupling to the MKID antennas is achieved by using a large monolithic lens array of elliptical Si lenses mounted on the chip backside and aligned so that each MKID antenna is located at the focus of an individual lens \citep{Filipovic1993}. The lens array is made commercially using laser ablation of high-resistivity Si ($\mathrm{\rho >5k \Omega cm}$) and equipped with a $\lambda$/4 anti-reflection coating made from parylene-C \citep{Ji2000}. The lens-antenna design is optimised for detection in a 170-GHz band centered around 850 GHz. 
%==========================================================================================================================
\begin{figure*}
\centering
\centering
\includegraphics[width=1\textwidth]{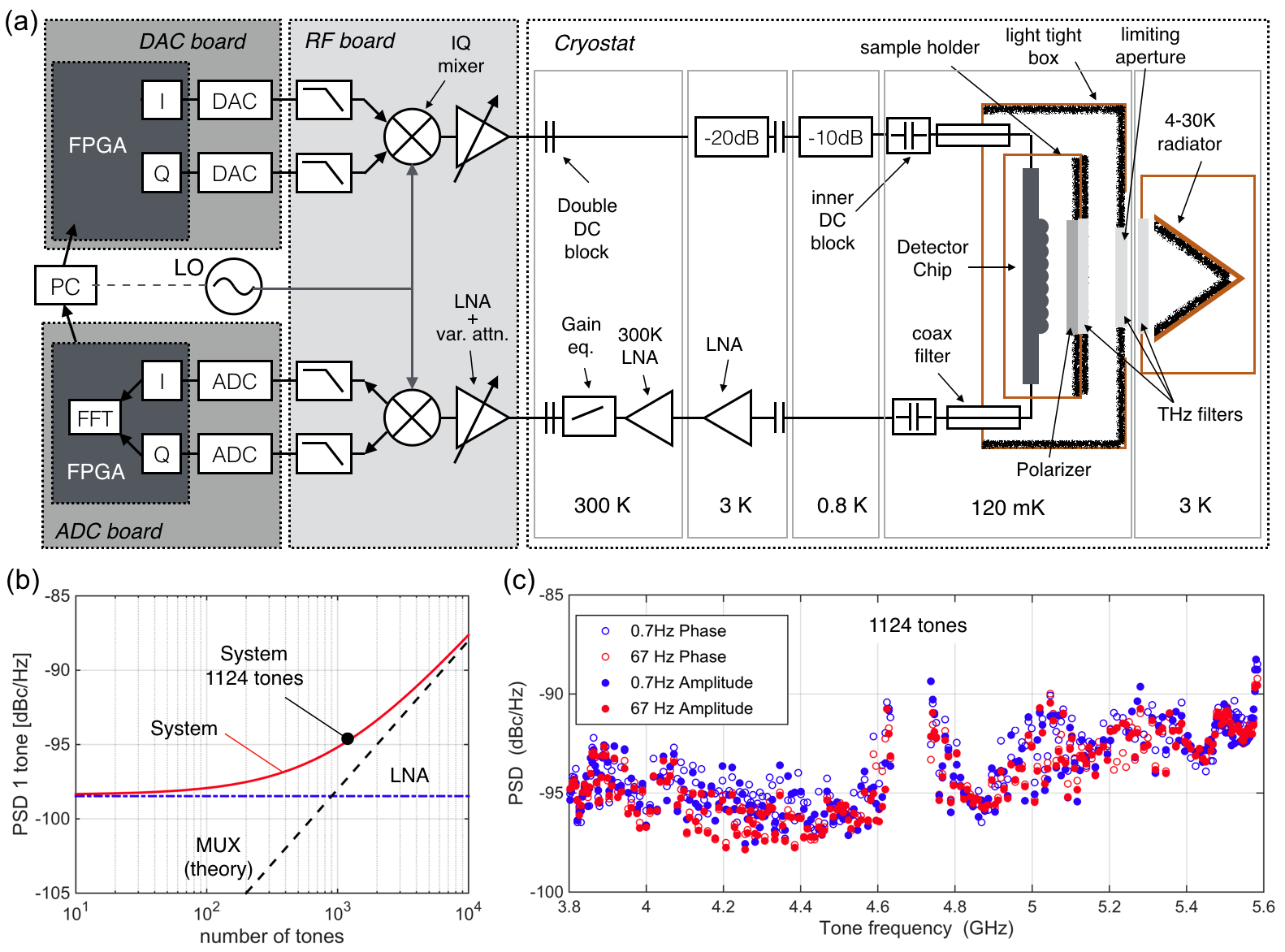}
\caption{(a) Schematic diagram of the experimental system, showing the digital and analogue sections of the readout system, the signal chain for the readout input and output lines and all components used. Note that the chip is mounted inside a light-tight box with coaxial feed-though filters for the readout signals; the thick black structure represents infrared absorber inside the sample holder and light-tight box to minimise stray light. (b) Calculated performance of the readout system, expressed in power spectral density of an individual readout tone, as a function of the number of readout tones used. Note the linear decrease in performance with increasing number of tones. (c) Measured performance of the readout system with 1124 tones. 874 tones are placed on MKIDs and 250 tones are placed in between MKID resonances. The figure shows the performance of these 250 blind tones. The data are corrected for systematic noise contributions as explained in the text. The open shapes represent the phase noise, and the closed shapes represent the amplitude noise. The colour coding indicated in the legend gives the post-detection frequency at which the noise spectral density is evaluated. The measured performance is very similar in phase and amplitude and also independent of the post-detection frequency down to 0.7 Hz. We do see a small frequency dependence and the deterioration of the readout system near the LO, which is placed at 4.685 GHz. }
\label{Fig:Setup}
\end{figure*}
%==========================================================================================================================
All MKIDs in the array are coupled to a single feedline as indicated in Fig. \ref{Fig:1}(a,b). The feedline is a CPW with a central linewidth = 20 $\mu$m and a gap =  10 $\mu$m equipped with bond-pads at either end for connecting the chip to the readout circuit. To prevent excess inter-pixel crosstalk we need to connect the two ground planes of the feedline \citep{Yates2014}, which is achieved by placing two aluminium bridges in between each pair of MKIDs, isolated from the central line by a polyimide stub. The polyimide stub is created by spin-coating, baking and a photolithographic step to define the stub locations. A three-hour 250$^{\circ}$ C cure under nitrogen atmosphere is done to make the polyimide stubs chemically resistant to further processing steps.

The spatial encoding of the MKID resonant frequencies on the array is based on the scheme presented in \citet{Yates2014}: F$_0$ = F$\mathrm{_c}$ + {\bf M} $\times$ dF, with dF=1.6649 MHz,  F$\mathrm{_c}$ = 5 GHz and {\bf M} a 2-D matrix constructed from a spiral 1-D array with interleaving indices and an index gap in its centre as shown by the insert in Fig. \ref{Fig:1}(a). The result is that nearest-frequency MKIDs are separated by one extra detector, but never more. This separation is enough to mitigate EM cross coupling \citep{Yates2014}, but is kept small to be less sensitive to thickness variations of the NbTiN film \citep{Thoen2016}.

A key parameter in the design is the bandwidth of each resonator, which is defined by the coupling structure and denoted as the coupling Q factor Qc. We design the resonators to have  Qc = $\mathrm{1\times10^5}$, which is a compromise between high dynamic range (requiring a lower {Q factor) and a low probability of overlapping resonance features, resulting in  a better pixel yield, which requires a high {Q factor. The rule of thumb, obtained using statistical simulations of the resonator resonant frequency scatter due to fabrication limits, is that the number of pixels per octave of bandwidth $\leq$  Qc}/50. Hence Qc = $\mathrm{1\times10^5}$ for 1000 pixels in the 4-6 GHz readout band. Further increasing the pixel density results in pixel loss due to overlapping resonators.

The detector chip is equipped with a mesh structure made from 50-nm sub-stoichiometric TiN  \citep{Coiffard2016} located on the backside of the chip  as indicated in Fig. \ref{Fig:1}(e), which has a sheet resistance of 33 $\Omega$ and a critical temperature between 0.6 and 1.4 K with the lowest critical temperature in the chip center. The mesh has a circular aperture with R = 0.55 mm to allow unhindered radiation coupling from the lens to the antenna (Fig. \ref{Fig:1}(a,e)). The functions of this mesh are i) to absorb radiation not coupled to the antenna but scattered into the chip, thus preventing this stray radiation from coupling to other pixels \citep{Yates2017}, and ii) to absorb high energy phonons resulting from cosmic ray interactions as demonstrated by \cite{Monfardini2016}. The mesh design is optimised for radiation absorption in the 850-GHz frequency range, and to be fully transparent at the 4-6 GHz MKID frequency.  

Fig. \ref{Fig:1}(c) shows the chip and Si lens array mounted in the holder. The holder aperture has a diameter that is smaller than the chip due to the size of the available FIR filters in our measurement setup. At the right side of the holder there are two SMA connectors which are connected to the feedline using wire bonding. These connectors form the interface between the sample and the readout system shown in Fig. \ref{Fig:Setup}.

\section{Experimental system} 
\label{Sec:ExpSys}
The experimental system consists of an in-house built readout system, coaxial cables and amplifiers to connect the readout system to the detector array, and a commercial adiabatic demagnetisation (ADR) cooler. The cooler has 50-K and 3-K temperature stages cooled by a continuous pulse-tube cooler and additional 0.8-K and 120-mK stages cooled by two independent ADRs. The experimental system is schematically depicted in Fig. \ref{Fig:Setup} and we discuss it in detail below.

\subsection{Readout system}%%%
We use frequency division multiplexing to read-out all the 961 pixels in the array simultaneously. For this we use the digital-analogue readout system described in \citet{Rantwijk2016}, which is capable of reading out up to 4000 pixels in a 2-GHz bandwidth centred around a local oscillator (LO) frequency in the range of 4.5 - 7.5 GHz. This system is unique in its large readout bandwidth of 2 GHz and higher operating frequency compared to other, similar systems \citep{Bourrion2011,McHugh2012}, which makes it ideal for our MKID design. The readout normally operates at a data rate of 159 samples per second. At this setting the tone frequencies are limited to multiples of 3.8 kHz. It is also possible to use a faster sampling rate of 1.2 kHz, but this limits the tone frequencies to multiples of 30.5 kHz. The readout signal is fed into the cryostat by standard flexible coax cables with SMA connectors. We use a double DC block at the cryostat input to prevent ground loops. Inside the cryostat we use 2.19-mm diameter Cu:Ni semi-rigid coax cables with an Ag cladding on the central conductor to bring the signal from 300 K to 4 K, and we use a -20 dB attenuator at 4 K and a -10 dB attenuator at 0.8 K to reduce 300 K thermal noise to values corresponding to below 1 K. We have carefully tested many attenuators and found that the api-INMET attenuators work well at sub-K temperatures. At temperatures below 2.7 K we use 0.86 mm diameter  NbTi coax cables, which provide lossless signal transfer and adequate thermal isolation. To connect to the chip we use a single DC block and a dedicated low-pass feed-through filter that is part of the light-tight box surrounding the sample holder. \citep{Baselmans2012}. The signal passes through the chip where the MKIDs modify the amplitude and phase of the readout tones, and is transferred back to the 3-K stage using an identical filter, DC block and NbTi cable. At 2.7 K we use a dual stage Yebes 4-8 GHz low noise amplifier (LNA) with a noise temperature of 5 K to amplify the readout signal \citep{Diez2003}. The bias of the second stage of the LNA is slightly higher than nominal to increase the dynamic range of the LNA to a 1 dB input compression point of -31 dBm which is essential to prevent clipping of the LNA by the peak envelope power (PEP) of the readout signal: 
\begin{equation}
 PEP(dBm) \sim P_{tone} + C + 10\log{(n_{tones})},
\end{equation}
with $P_{tone}$ the tone power in dBm, C the Crest factor, which is $\sim$ 14 dB for random phases of the tone signals and $n_{tones}$ the number of readout tones. We have found that a PEP at the LNA input up to 6 dB below the 1-dB compression point does not create significant harmonics in the spectrum of the readout signal, so we use PEP $<$ -37 dBm. This value allows up to $\sim 10^4$ readout tones of -92 dBm RMS power. Further amplification is implemented at room temperature at the top of the cryostat and inside the analogue board of the readout electronics. We use a positive slope gain equaliser of 2 dB/GHz to create a flat system transmission, compensating for frequency dependent cable losses; this guarantees a frequency independent tone power at the ADC. 
%==========================================================================================================================
\begin{figure*}
\centering
\includegraphics[width=1\textwidth]{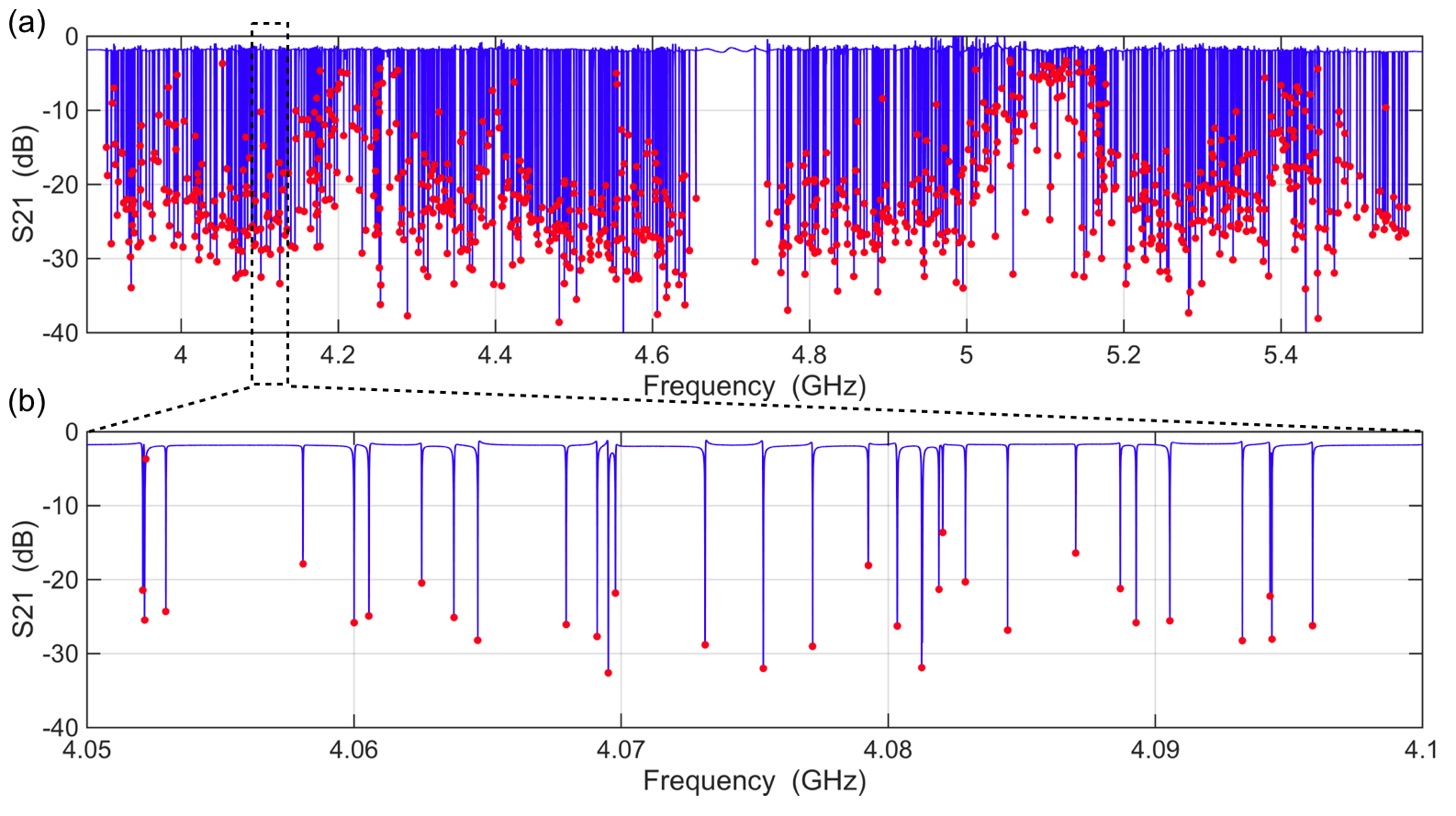}
\caption{(a) Frequency sweep of the array, with each dip corresponding to a different MKID pixel. The response in the figure is relative to a calibration performed at 0.8K, which effectively removes a $\sim$3dB ripple due to the setup cabling.  (b) Zoom of a section of panel (a), showing the relative bandwidth of the resonators and the scatter in frequency of the resonators, mainly due to thickness variations in the NbTiN.} 
\label{Fig:Fsweep}
\end{figure*}
%==========================================================================================================================

In the experiments we place one readout tone close to the resonance of each resonator. Importantly we place also 250 readout tones at frequencies \emph{not} affected by a resonance feature. Referring to Fig. \ref{Fig:Setup}(b) adding 250 tones gives a $\sim$ 1dB deterioration of the system noise that does not affect the measured phase noise of the resonators. We refer to these readout tones as blind tones and use them to correct for systematic noise in the readout chain: For each sample (i.e. each data point in the time domain for all readout and blind tones) we spline-fit the measured tone frequency dependence of the phase and amplitude values of all blind tones. This gives us a function that describes the phase delay and amplitude transmission at any arbitrary frequency within the range of the blind tones. We divide each blind tone and each readout tone by the corresponding interpolated complex value. This removes the phase delay due to the cables, amplitude ripple and any time-dependent changes of these quantities. 

The blind tones also allow us to measure the intrinsic performance of the readout system, shown in Fig. \ref{Fig:Setup}(c): We take several minutes of data with both blind tones and MKID readout tones. We calculate the power spectral density (PSD) of the noise of each blind tone after correcting for systematics as described before. We plot the averaged PSD around two post-detection frequencies (0.7 Hz and 67 Hz) in Fig. \ref{Fig:Setup}(c). We observe a noise performance given by a power spectral density PSD$\sim$-95 dBc/Hz in the two frequency bands plotted showing that the noise spectra are white, in agreement with \citet{Rantwijk2016}. This value can be compared to a detailed prediction of the system performance that takes into account all components of the analogue up- and downconverter boards and the measured performance of the digital system, which is shown in Fig. \ref{Fig:Setup}(b). We see that the readout system performs as predicted and that, at 1124 readout tones, the overall performance is equally limited by the digital system and the analogue system which has an effective noise temperature of 6.5 K at the MKID level. This value is dominated by the noise temperature of the Yebes LNA (5 K). 

\subsection{sub-mm system}

The detector chip is mounted inside a light-tight holder which is in turn mounted within another light-tight box. Both the sample holder and the light-tight box have a significant fraction of their inner surfaces coated with EPOTEK 920 epoxy mixed with 3\% by weight carbon black in which we embed 1-mm RMS diameter grains of SiC \citep{Baselmans2012}. This box-inside-a-box strategy guarantees low enough stray light levels; for more details see \citet{Visser2014, Baselmans2012,Barends2011}. 

Note that the set-up described here differs from the one described by \citet{Visser2014} only in the frequency band of the sub-mm filters between the chip and the calibration source. The calibration source is a black body radiator made from a copper cone coated on the inside with the same strongly absorbing material used for the absorbers in the light-tight box and sample holder. The cone is weakly thermally coupled to its surrounding copper box, which itself is thermally anchored to the 2.7-K stage of the cooler. Resistive heaters and a thermometer allow us to control the radiator temperature between 2.7 K and 30 K. Radiation is coupled to the chip through three stacks of sub-mm filters that define a passband of 825 GHz - 905 GHz: Together they have a measured out-of-band rejection of -50 dB, which is especially important at low black body temperatures ($\mathrm{T_{BB} <}$ 5K) where the vast majority of the black body spectral emission is at frequencies below 850 GHz. The throughput-limiting aperture to the array is the opening of the 120-mK cold box, with a diameter of 22 mm at a distance of 18 mm from the chip. There is an additional aperture on the sample holder clearly visible in Fig. \ref{Fig:1}(c). The consequence of these apertures is that only the central $\sim$ 40 pixels of the array are illuminated, and that the edge pixels of those will have only a partial coupling to the calibrator due to the throughput-limiting aperture in the cold box.
 
\section{Experiments}
\label{Sec:Exp}

\subsection{Resonator characterisation}
To characterise the detector array we stabilise the chip at a temperature of 120 mK and operate the black body with zero heater current, resulting in T$_{BB} \sim$  2.7 K. We refer to this condition as 'cold and dark' in the remainder of the text. We use a commercial vector network analyser instead of the multiplexed readout  in Fig. \ref{Fig:Setup}(a) to measure the forward scattering parameter S21 of the system as a function of frequency; the result is shown in Fig. \ref{Fig:Fsweep}. We observe a 'forest' of resonance features, each one corresponding to an individual MKID and a frequency-independent transmission where no resonances are present. The resonances occupy a frequency range from 3.9 - 5.55 GHz with a small gap (by design) in the centre of the band which is used to place the LO of the readout electronics. The frequency range is 5\% lower than the design due to a slightly higher kinetic inductance in the MKIDs, which is of no consequence for the pixel performance. Using an algorithm based on the double derivative of the presented data to identify all resonators, we find 907 resonators out of 961, i.e. 94\%. We fit to all the resonance features a Lorentzian function to extract the Q factor and depth of each resonance, from which we can deduce the coupling Q factor Qc and Qi the Q factor describing all other losses in the MKID resonator: Q$^{-1}$ = Qc$^{-1}$+Qi$^{-1}$. We find that $\mathrm{<Qc>}$ = 1.3$\pm$0.4$\times10^5$, close to the design value and that $\mathrm{<Qi>}$=1.3$\pm$0.9$\times10^6$,  and  $\mathrm{Qi>}$  5$\times\mathrm{10^5}$  for 715 devices, i.e. at cold and dark conditions the resonator Q is dominated by Qc. Several of these scans were performed to determine the optimum readout power of the detectors. This is the maximum power for which the MKID resonance features have no signs of asymmetry. We observe that -92 dBm readout power at the MKID chip allows all resonances to be read-out; at -86 dBm more than half of the MKIDs are overdriven: they are asymmetric and produce very significant excess noise.

\subsection{Experimental methodology to measure the NEP}
\label{Sec:Meth}

To measure the detector optical efficiency and sensitivity we use the method developed by \citet{Janssen2013}: We measure the detector NEP under background-limited conditions (i.e. at sufficiently high temperature of the black body calibrator) and compare the result to a theoretical calculation of the photon noise limited sensitivity. This analysis allows a direct measurement of the optical efficiency and requires an analysis of the spectral shape of the noise to ensure background limited operation of the MKID. In this section we discuss this method in detail, the results are given in Sections \ref{section:Efficiency} and \ref{section:Sensitivity}.      
 %==========================================================================================
\begin{figure*}
\centering
\includegraphics[width=1\textwidth]{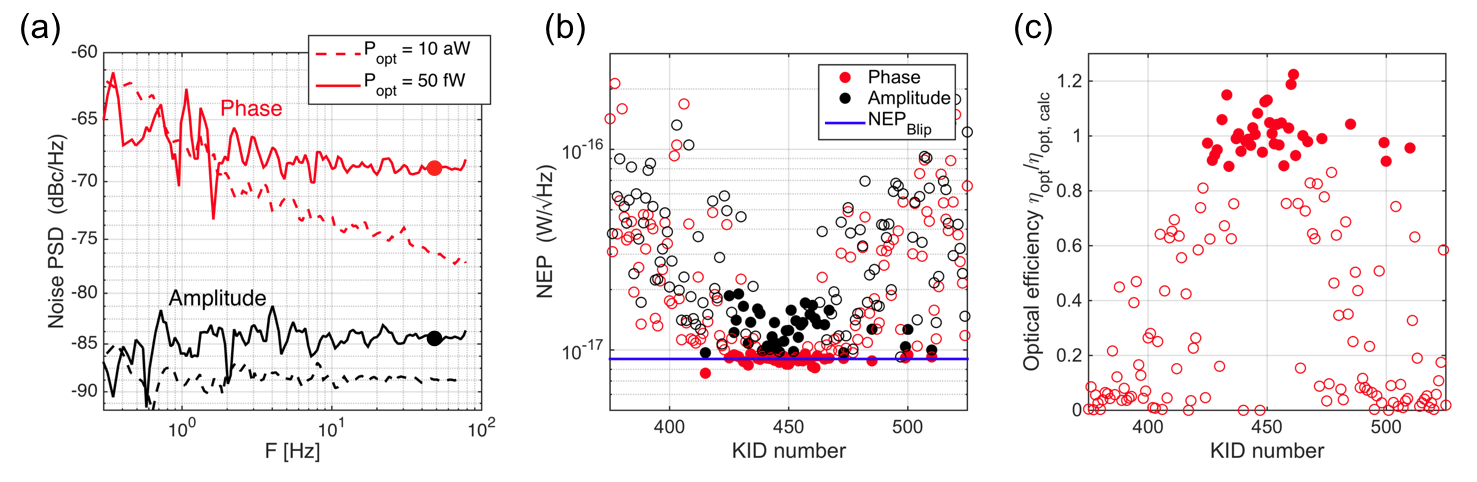}
\caption{(a) Measured phase and amplitude noise of a representative MKID at 10 aW of absorbed power per pixel (cold and dark conditions, 2.7-K calibrator) and at 50 fW absorbed power (9-K calibrator). Note that the spectra are taken from the signal relative to the circle traced by the frequency sweep around the resonator; see \citet{Gao2008b} for details.  (b) Measured NEP using phase and amplitude readout at a post-detection frequency between 60 and 80 Hz at a black body temperature of 9 K. The NEP calculation is done using Eqn. \ref{Eqn:OptEf} with the calculated obtained optical efficiency $\mathrm{\eta_{opt,calc}}$. The blue line is the background limited $\mathrm{NEP_{ph}}$ obtained using \ref{Eqn:Popt} and Eqn. \ref{Eqn:OptEf}  with $\mathrm{\eta_{opt,calc}}$. (c) Experimental optical efficiency relative to the calculated value, obtained from the difference between the measured phase NEP and the background limit, both shown in panel (b). In both panels (b) and (c) the solid symbols represent the detectors that are fully illuminated by the calibration source, the open symbols represent the detectors that are not, or partly, illuminated.}
\label{Fig:OpticalEf}
\end{figure*}
%============================================================================================
                   
We use the multiplexed readout in the configuration shown in Fig. \ref{Fig:Setup} and operate it in its standard configuration with a tone placement in multiples of 3.8 kHz and a data rate of 159 samples/sec. We use $\mathrm{F_{LO} = 4.685}$ GHz, in the centre of the empty frequency region in Fig. \ref{Fig:Fsweep}(a). We stabilise the array temperature at 120 mK and put the array under the desired FIR loading conditions by stabilising the calibrator temperature at the appropriate value. We will use a calibrator temperature of 9 K and prove that this represents background limited operation of the MKIDs. We also measure under 'cold and dark' conditions where the radiated power is negligible. To initialise the measurement sequence we perform a `wide frequency sweep' using 1000 evenly spaced readout tones at -92 dBm of power per tone by sweeping the LO frequency. The result of this measurement is identical to the data presented in Fig. \ref{Fig:Fsweep}(a). The system software finds all resonance features and places readout tones as closely as possible to the central frequency of each resonance. Given an MKID bandwidth of $\sim$ 50 kHz, we can place the readout tone with a resolution of $\sim$BW/13. In total 874 tones are placed, marginally fewer than the 907 resonance we found. This because the readout system software rejects tones close to the LO frequency and band edges. We also place 250 blind tones to correct for systematics as explained in Section \ref{Sec:ExpSys}.

To start a NEP measurement we perform a `local frequency sweep': we sweep the LO frequency over a range of $\pm$0.25 MHz around $F_{LO} = 4.685$ GHz with 1-kHz steps; this corresponds to a scan width of $\sim$ 10 resonator bandwidths and 50 frequency points per bandwidth. These results are saved so as to be able to measure the phase $\theta$ and amplitude signal $A$ of each resonator relative to the circle traced in the complex plane by the local frequency sweep. This method, pioneered by \citet{Gao2007} allows the use of phase and amplitude readout in which only the phase readout is affected by device intrinsic TLS noise \citep{Gao2008b, Gao2008}. To obtain the device sensitivity using phase readout we measure the phase noise power spectral density (PSD) $S_{\theta}$, the phase response of the device to a small change in absorbed power $\mathrm{\delta\theta/{\delta P_{opt}}}$, and response time, given in our case by the quasiparticle lifetime $\tau_{qp}$. From these measurements the experimental NEP can be obtained using \citet{Visser2014}
\begin{equation}
NEP_{\theta} = \sqrt{S_{\theta}} \left(\frac{\delta\theta}{\delta P_{opt}}\right)^{-1}\sqrt{1+\omega^2\tau_{qp}^2}
\label{Eqn:OptNEP}
\end{equation}
We determine the amplitude NEP in the same way by using the amplitude noise $\mathrm{S_{A}}$ and amplitude responsivity $\mathrm{\frac{\delta A}{\delta P_{opt}}}$. 

To characterise the noise we take several minutes of data with $F_{LO} = 4.685$ GHz and correct all data for systematics using blind tone correction as explained before. Subsequently we use a linear transformation to bring, for each MKID, the coordinate system to the centre of the MKID resonance circle. As a last step we remove glitches in the time domain data that are caused by cosmic ray interactions using the procedure explained in section \ref{Section:Cosmic rays}. We calculate the phase and amplitude noise power spectral density $\mathrm{S_{\theta}}$ and $\mathrm{S_A}$ from the de-glitched time domain data using a standard FFT routine. The spectra for a representative MKID are shown in Fig.  \ref{Fig:OpticalEf}(a). Under 'dark and cold conditions' (blue lines) the phase noise spectrum has a spectral dependence that can be best described by a combination of a 1/f and 1/$\sqrt{f}$ noise. This is typical for TLS noise \citep{Gao2008}. Under these conditions the amplitude noise spectrum is white and limited by the readout system noise. With the black body at 9 K, both the phase and amplitude noise spectra become white. Additional measurements with a single-tone readout show that the noise spectra both roll off with an identical time constant given by the quasiparticle lifetime and that the phase noise is limited by the MKID itself, the amplitude noise still has a small contribution from the readout system. These two observations prove that we reach background limited performance with phase readout with the calibrator at a temperature of 9 K (see \cite{Janssen2013} for more details). 

The MKID phase and amplitude responsivity, $\delta\theta/\delta P_{opt}$ and  $\delta A/\delta P_{opt}$, are obtained by a linear fit to the de-glitched phase and amplitude response of the MKID to a change in absorbed power $P_{opt}$. We create a small change in $P_{opt}$ by varying the temperature of the black body calibration source. Note that the measurement is done both for increasing and decreasing black body temperature to eliminate hysteretic effects. We convert the black body temperature into the power absorbed in the detector $P_{opt}$ by using the method described in \citet{Janssen2013}: 
\begin{equation}
P_{opt}=\eta_{opt}\int_{\nu}  \frac {1}  {2}  F_\nu B_\nu (T_{BB})   \lambda^2 d\nu.
\label{Eqn:Popt}
\end{equation}
Here $B_\nu$ is the radiator intensity, the factor 1/2 reflects the fact that we use a single polarisation, $\lambda^2=A\Omega$ is the total throughput of the detector, and $F_\nu$ is the measured transmission of our FIR filters. The optical coupling $\eta_{opt} = \eta_{rad} \times \eta_{SO} $ describes the fraction of the power emitted from the calibrator that is absorbed in the aluminium strip of the MKID. Here $\eta_{SO}$ represents the spillover between the detector beam and the limiting aperture to the calibrator and the radiation efficiency $\eta_{rad}$ is the fraction of the power falling on the lens that is absorbed in the MKID aluminium strip. Using detailed   CST (Computer Simulation Technology) simulations of the lens-antenna beam in combination with the setup geometry we can calculate from the setup geometry and the simulated lens-antenna beam pattern that $\eta_{opt,calc}$ = 0.61, with $\eta_{SO,calc}$ = 0.82 and $\eta_{rad,calc}$ = 0.74. 

We obtain the time constant of the detector, which under cold and dark conditions is given by $\tau_{qp}$, by performing a single exponential fit to the tail of cosmic ray pulses as discussed in Section \ref{Section:Cosmic rays} and shown in Fig. \ref{Fig:CR}. We obtain a mean value $<\tau_{qp}>$ = 1.48 ms under cold and dark conditions. We use these average values for all MKIDs because we cannot obtain a good lifetime estimate for each individual resonator due to limitations in readout tone placement accuracy when using the 1.2 kHz data rate option, which is needed to measure a msec timescale. For a 9 K calibrator temperature we obtain $<\tau_{qp}>$ = 0.3 ms using a single tone readout system for the six central pixels of the array.

Under the condition that the MKID performance is background limited its NEP should be identical to the theoretical photon noise limited NEP of an MKID which is given by \citep{Flanigan2017,Zmuidzinas2012}:
\begin{equation}
NEP_{Blip}^2 = 2P_{opt}h\nu(1+\eta_{opt}F_{\nu}B)+4\Delta P_{opt}/\eta_{pb}
\label{Eqn:OptEf}
\end{equation}
with $\Delta$ = 0.1957 meV, obtained from the measured Tc = 1.29 K of our aluminium film and the BCS relation $\mathrm{\Delta=1.76 k_BTc}$ with $\mathrm{k_B}$ Boltzmann's constant. Furthermore, h is Planck's constant, $\nu$ is the detecting frequency (850 GHz), B the photon bunching term and $\mathrm{\eta_{pb}}$ = 0.4 is the quasiparticle creation efficiency, which is modified from the more conventional value of 0.57 using the results from \citet{Guruswamy2015} and the parameters of our aluminium film: $h\nu/\Delta \sim 20$ and $\tau_l/\tau_0^{\phi} \sim 0.5$. Note that the last term in the equation describes the NEP contribution due to quasiparticle recombination. This recombination term increases the NEP$\mathrm{_{Blip}}$ by a factor 1.09 for the range of black body powers used in our experiment. Looking at Eqn. \ref{Eqn:OptNEP} and Eqn. \ref{Eqn:OptEf} we see that they are identical only for a unique value of $\mathrm{P_{opt}}$, i.e. for a unique value of $\mathrm{\eta_{opt}}$. Hence a measurement of the NEP using Eqn. \ref{Eqn:OptNEP} and comparing it to the calculated Eqn. \ref{Eqn:OptEf} allows for an experimental verification of the optical efficiency of the detector. 

\subsection{Optical Efficiency}\label{section:Efficiency}

The methodology explained in the previous section allows us to find the experimental value of the optical efficiency from a comparison between the experimentally obtained NEP and the theoretical, background limited, $NEP_{ph}$. The solid blue line in Fig. \ref{Fig:OpticalEf}(b) represents $NEP_{ph}$ calculated using Eqn. \ref{Eqn:Popt} and \ref{Eqn:OptEf} with the $\eta_{opt,calc}$ = 0.61 obtained from our detector model and the geometry of the experimental setup. The dots are the measured NEP values obtained using Eqn. \ref{Eqn:OptNEP}. The noise level used for the calculation is mean noise level in the range 60-70 Hz as indicated in Fig.\ref{Fig:OpticalEf}(a). We see that the phase NEP has less spread and has a lower mean value that than the amplitude NEP. This is caused by the amplitude NEP having a small noise contribution from the readout system: i.e. only the phase readout gives a background-limited detector performance. We use Eqn.\ref{Eqn:OptEf} together with the measured phase NEP to obtain the optical efficiency from the experiments, $\eta_{opt}$. In Fig. \ref{Fig:OpticalEf}(b) we plot $\eta_{opt}/\eta_{opt,calc}$ and find that it is $1 \pm 0.1$ for the central pixels of the array. The efficiency is reduced for the non-central pixels due to (partial) obstruction of the detector beam by the apertures in the system. 
%The slight reduction in coupling efficiency compared to the theoretical value is probably due to the glue gap between the lens array and the chip. Simulations show that a 10$\mathrm{\mum}$ gap is sufficient to reduce $\mathrm{\eta_{rad}}$ by 8\% compared to a 5$\mathrm{\mum}$ gap we used in the calculation. Hence we conclude that the experimental optical efficiency $\eta_{opt}$ = 0.9 $\times\eta_{opt,calc}$ = 0.56. 
%==========================================================================================================================
\begin{figure}
\centering
\includegraphics[width=1\linewidth]{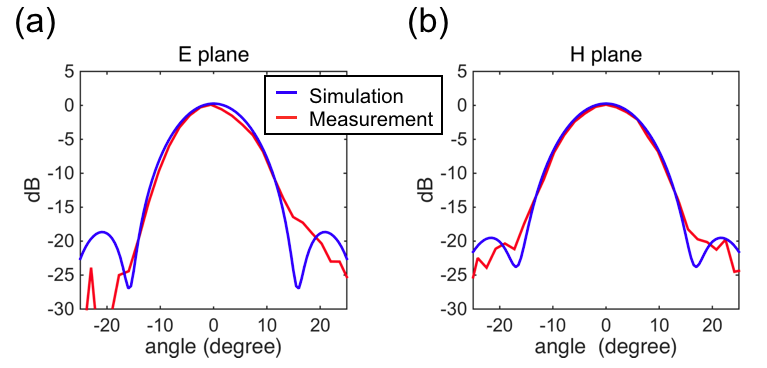}
\caption{Measured and calculated beam pattern of a single detector on a test chip with identical design, and fabricated on the same wafer, as the array presented here. An excellent agreement for the main beam is observed. Panel (a) shows the results in the E plane (cross section perpendicular to the antenna slots) and (b) in the H-plane (cross section parallel with the antenna slots).} 
\label{Fig:BP}
\end{figure}
%==========================================================================================================================
%===========================================================================================================
\begin{figure}
\centering
\includegraphics[width=1\linewidth]{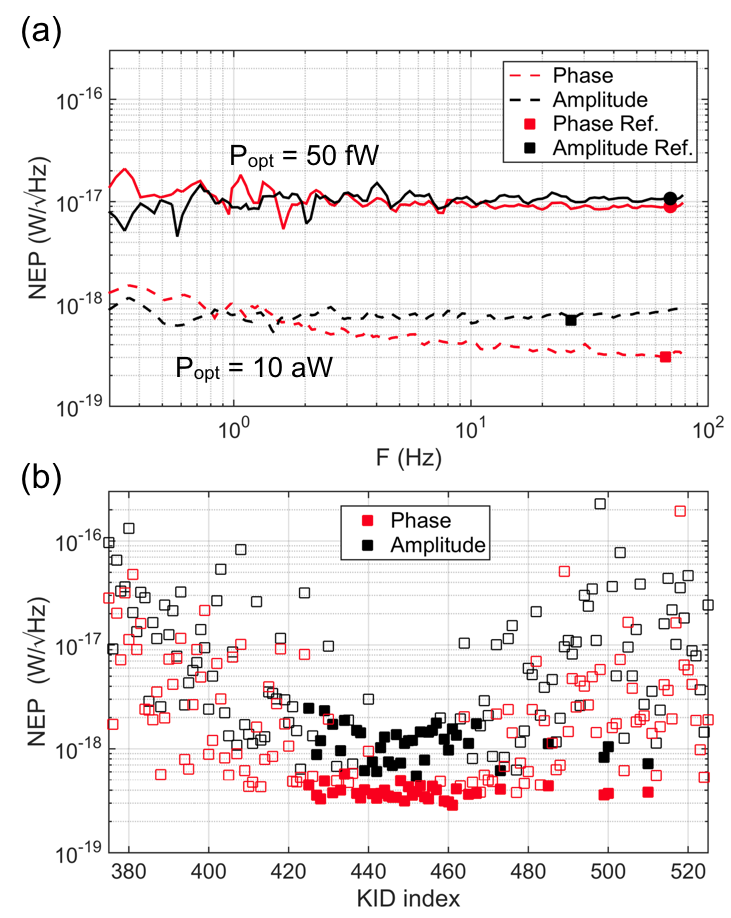}
\caption{(a) Measured detector NEP referred to the power absorbed in the pixel at 10 aW (bottom lines), which represents the performance limit of the detectors, and 50 fW (upper lines). Note the white noise spectra and identical sensitivity using amplitude or phase readout in this case. The dots represent the minimum of the NEP curve shown in panel (a) (10 aW case) and the NEP in a  60-80 Hz  band for the 50 fW case. (b) Minimum NEP, using phase- and amplitude readout at 10 aW loading. The NEP calculation is done using Eqn. \ref{Eqn:OptEf} with the experimentally obtained optical efficiency. The solid symbols represent the detectors that are fully illuminated by the calibration source.}
\label{Fig:OptNEPlim}
\end{figure}
%======================================================================================================

As stated before the optical efficiency is the product of the spillover efficiency, determined by the beam shape and apertures in our system, and the radiation efficiency, determined by the lens-antenna assembly. To be able to distinguish between the two effects we have measured the lens-antenna beam pattern using a small test chip with an identical MKID and an identical lens design. The experiments and results are presented in \citet{Ferrari2017}; for completeness we reproduce here the measured and calculated patterns in the E plane (E-field perpendicular to the slots) and in the H plane in Fig. \ref{Fig:BP}. We observe an excellent agreement between the measurements and calculations for the main beam. Given the fact that the beam pattern matches the measurements presented in Fig. \ref{Fig:BP} very well, we conclude that the calculated value of the spillover, which is determined by the beam pattern, correctly represents the experiment. As a result $\mathrm{\eta_{SO}=\eta_{SO,calc}}$ = 0.82. Given that we found before that $\mathrm{\eta_{opt}=\eta_{opt,calc}}$ we can conclude that  $\mathrm{\eta_{rad}=\eta_{rad,calc} = 0.74}$. Additionally the agreement between the measured and simulated antenna beam pattern allows us also to calculate the taper efficiency from the simulated beam pattern. We find $\eta_{\mathrm{tap}}$ = 0.78. 

%============================================================================================
\begin{figure*}
	%\begin{minipage}{1\textwidth}
	\centering
	\includegraphics[width=1\textwidth,keepaspectratio]{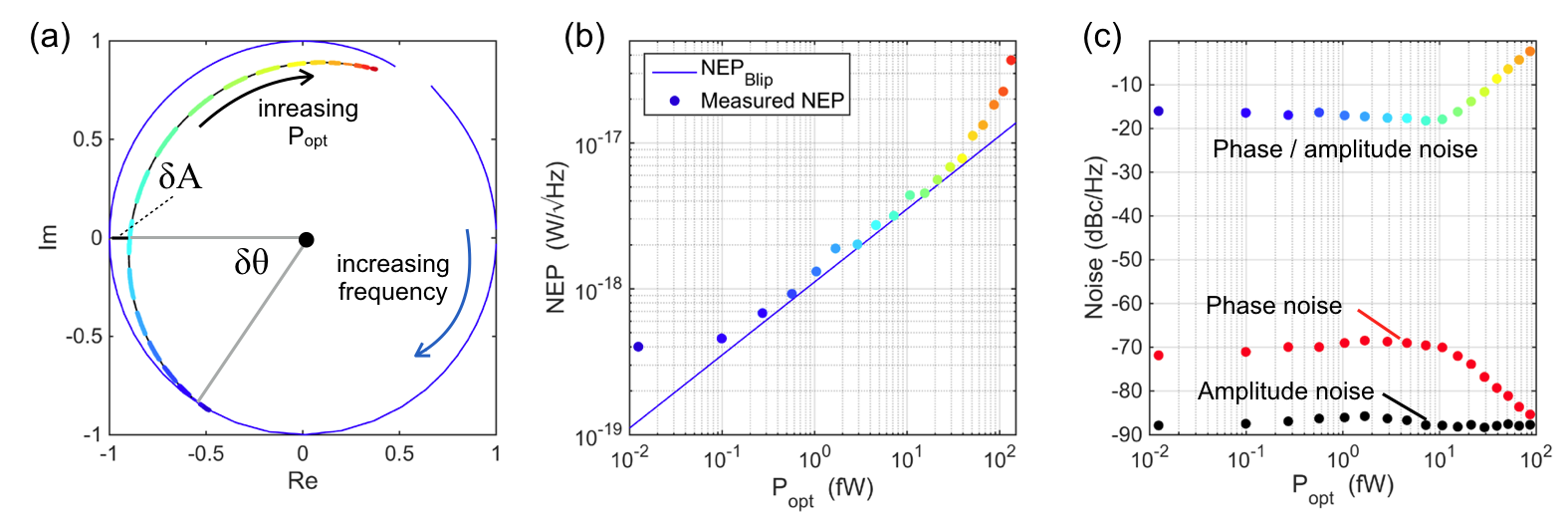}
	%\end{minipage}
	\caption{ (a) The MKID local frequency sweep results in a circle in the complex plane, shown by the blue circle. The line inside the circle is the trace of the MKID response from an increase in absorbed power from $<$ 10 aW to 200 fW with a readout tone placed 10 kHz below the MKID resonance frequency. The data are over-plotted with the sections used to calculate the NEP. (b) NEP as a function of loading power with a constant LO frequency at a post-detection frequency of 60-80 Hz. We see that the measured NEP approaches the $\mathrm{NEP_{Blip}}$ for powers exceeding 100 aW up to 40 fW. (c) The mean value in a 60-80 Hz post detection frequency band of the phase noise power spectral density, amplitude noise power spectral density and the ratio of the two. We see that at low power the phase noise exceeds the amplitude noise. At the highest power we see that the phase and amplitude noise become similar due to a sharp drop in the phase noise level. This drop is caused by the MKID responsivity being reduced by the resonator moving too far off the readout tone. The power level where the phase noise approaches the amplitude noise, which is the same power level at which the NEP starts to deviate from the theoretical prediction in panel (b), defines the maximum instantaneous source power that the device can measure. }
\label{Fig:DynRange}
\end{figure*}
%==========================================================================================
\subsection{Limiting Sensitivity}\label{section:Sensitivity} 

Now that we have an experimental measurement of the optical efficiency $\mathrm{\eta_{opt}}$ we can use it in combination with Eqn. \ref{Eqn:OptNEP} to calculate the limiting sensitivity to FIR/sub-mm radiation of the central pixels of the array. This is done under cold and dark conditions. The results are shown by the blue lines at the bottom of Fig. \ref{Fig:OptNEPlim}(a) for a representative pixel as a function of the post-detection frequency. We take the minimum of these curves , as indicated by the dots, for all pixels of the array and plot in Fig. \ref{Fig:OptNEPlim}(b) the detector NEP for the MKIDs that are illuminated by the calibration source. We observe that the phase readout NEP has a lowest value of $NEP \sim 3\times10^{-19}W/\sqrt{Hz}$, the spectrum is non-white due to TLS noise contributions from the detector as discussed before. The amplitude NEP has a lowest value of $NEP = 6\times10^{-19}W/\sqrt{Hz}$ for the best pixels, with a white spectrum but as before with more scatter in NEP between the pixels. The higher value and the scatter of the NEP is due to excess noise from the readout system, since the MKID output photon noise is much lower than the phase noise, \citep[see ][]{Janssen2013,Rantwijk2016}. For both amplitude and phase readout we see in Fig. \ref{Fig:OptNEPlim}(b) that the NEP is similar for all central pixels, but increases at lower and higher MKID index due to partial (or complete) obscuration between the radiator and the detectors (Fig. \ref{Fig:1}b and \ref{Fig:Setup}). 

To give an absolute measure of the performance of this detector system we must relate it to the imaging speed. The two figures of merit for this are: i) the intrinsic detector sensitivity, given in Fig. \ref{Fig:OptNEPlim}(b) and ii) the aperture efficiency, defined as $\mathrm{\eta_{ap} =\eta_{rad}\times\eta_{tap}=0.74\times0.78=0.58}$ where we take $\mathrm{\eta_{tap}}$ from the simulated antenna beam pattern and $\mathrm{\eta_{rad}=0.74}$ as discussed in Section \ref{section:Efficiency}. For a perfect single-mode system $\mathrm{\eta_{AE}\;\sim\; 0.8}$, the pixels of our array have a relative coupling efficiency of 73\%. The most important factor contributing to the reduction in coupling with respect to a perfect system is coupling loss due to the birefringence of the sapphire substrate. An antenna on a non-birefringent substrate such as silicon in combination with a better mounting technology for the lens array could mitigate both issues. 

\subsection{Dynamic range}\label{section:dynamic range}
To be able to map extended astronomical sources, which often have a large variation in surface brightness across the image,  the detectors of an imaging array must be able to cope with correspondingly large variations in $P_{opt}$ without saturation. When reading out the detector, increasing absorbed power shifts the MKID resonance frequency to lower values, as shown in Fig. \ref{Fig:1}(f). At some power level, the resonance will be shifted completely away from the readout tone, resulting in loss of detector response. This determines the instantaneous dynamic range of the system. To measure it we first perform a frequency sweep under dark and cold conditions to find the MKID resonant frequencies. Subsequently we take data while increasing the black body temperature up to 30 K, or $P_{opt}$ = 200 fW absorbed power. We use readout tones at $\mathrm{F_0 - 10kHz}$, representing a readout tone at a 0.25 bandwidth below the MKID resonance. The use of a lower frequency readout tone increases dynamic range and does not decrease the detector performance. The frequency sweep and detector response are shown for a representative detector in Fig. \ref{Fig:DynRange}(a), with the frequency sweep indicated by the blue circle, and the trace inside the circle representing the measured data at F$_0$-10kHz when increasing the black body power. The lowest FIR power corresponds to the low intersection of both curves. Here we observe an advantage of MKID phase readout: the detector phase response $\mathrm{\delta\theta}$ increases monotonically with absorbed power, whereas the MKID amplitude response $\mathrm{\delta A}$ is non-monotonic. 

To analyse the detector performance we divide the data into small sections of 2000 data points and, for each section, perform a linear fit to the phase response versus absorbed power to obtain $\delta \theta/\delta P_{opt}$. The number of data points is chosen to yield a linear response with sufficient signal-to-noise for a reliable fit. In Fig. \ref{Fig:DynRange}(a) the data sections are indicated by the dots on the MKID response curve. To get an estimate of the noise we subtract the fit from the data and calculate the power spectral density of this baseline corrected data. Using Eqn.\ref{Eqn:OptNEP} we calculate the phase readout limited NEP at a post-detection frequency of 80 Hz. In Fig. \ref{Fig:DynRange}(b) we show the measured phase NEP as a function of the absorbed power together with the calculated NEP assuming background limited detector performance, obtained using Eqn. \ref{Eqn:OptEf}. The result is almost identical to the result from \citet{Visser2014} obtained for a single pixel using amplitude readout: At the lowest power the MKID is detector-limited; increasing the radiation power we quickly approach a sensitivity very close to the background limit (given by the blue line), which is in agreement with the results in Section \ref{section:Efficiency}. At absorbed powers larger than 40 fW the NEP deteriorates faster than the theoretical curve because the MKID resonance feature has moved too far away from the readout tone. The result is a reduction in device responsivity, which reduces the MKID output noise power spectral density, and is clearly visible in Fig. \ref{Fig:DynRange}(c). This process, starting at 10 fW, results in the readout noise becoming significant and thereby increasing the NEP above the background limit. Note that the device output noise is at these powers purely given by the photon noise from the calibration source. The dynamic range of the detector is thus given by $P_{saturation}/NEP$ = $1\times 10^5$, and the maximum source power we can observe is 40 fW. The other detectors in the array that are illuminated by the black body calibrator give very similar results. This dynamic range is large enough for virtually all applications. If it would be possible to adjust the  frequency of the MKID readout signal, the dynamic range would be more than a factor ten larger.

\subsection{Dark Sensitivity}\label{section:darkNEP}
%==========================================================================================================================
\begin{figure}
\centering
\includegraphics[width=1\linewidth]{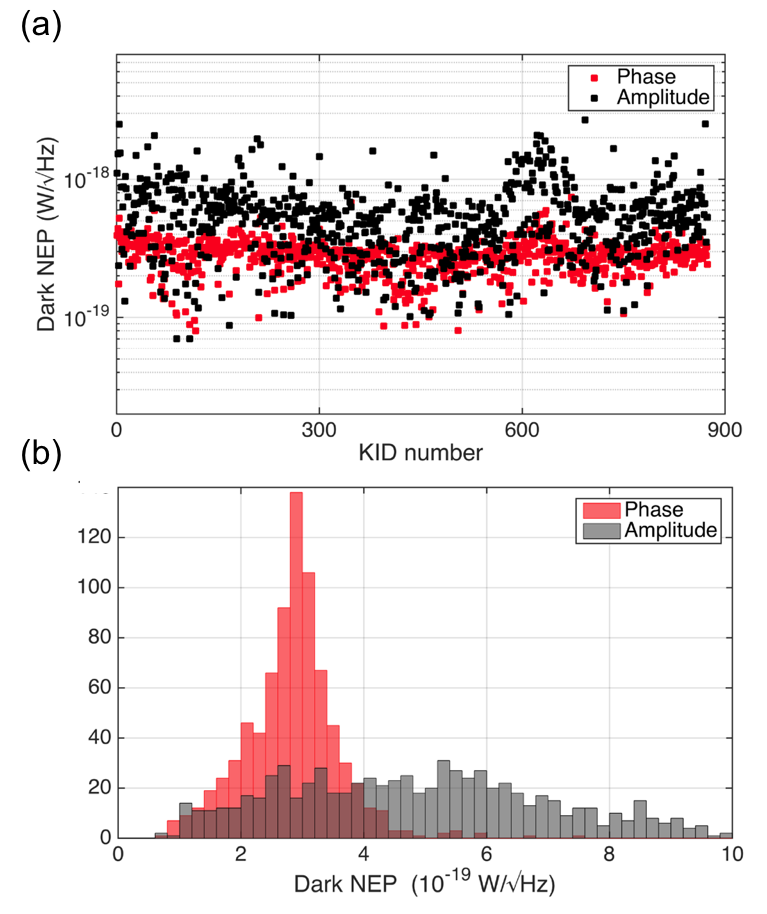}
\caption{(a) The minimum dark NEP for all MKIDs of the array, obtained by measuring the temperature response of the chip and the noise spectra at 120 mK; for details see the text. (b) Histogram of the data plotted in panel (a).} 
\label{Fig:DarkNEP}
\end{figure}
%==========================================================================================================================
The measurements presented so far have demonstrated the performance of the central pixels because of the limited aperture to the black body calibrator in our experimental system. To assess the performance of the entire array we have to measure parameters that are independent of the radiation coupling scheme. We therefore measure the `dark' NEP of the detectors using the method described in \citet{Baselmans2008}: We measure the response of the MKIDs to a change in chip temperature while keeping the radiator in dark conditions, i.e. $\mathrm{T_{BB}}$ = 2.7 K. Under these conditions the amount of quasiparticles in the aluminium $\mathrm{N_{qp}}$ can be calculate from the chip temperature, the volume of the aluminium section of the resonator and the energy gap \citep{Janssen2014b}. Rewriting $\mathrm{N_{qp}}$ in terms of FIR power allows the dark NEP to be calculated using Eqn.\ref{Eqn:OptNEP} by replacing the responsivity term $\delta\theta/\delta P_{opt}$  by the dark responsivity
\begin{equation}
\label{Eqn:NEPdark}
\frac{\delta \theta}{\delta P_{dark}} = \frac{\eta_{pb} \tau_{qp}}{\Delta}\frac{\delta \theta}{\delta N_{qp}(T)}
\end{equation}
or its equivalent for amplitude. Importantly \citet{Janssen2014b} have shown that the dark NEP is a good approximation for the NEP measured using a calibration source for the hybrid NbTiN-Al MKID design. 

We give, in Fig. \ref{Fig:DarkNEP} the dark NEP for both phase and amplitude readout. The dark NEP is  given by $\mathrm{NEP_{dark} = 2.8 \pm 0.8 \times10^{-19}\;\WHz}$ for phase readout. The scatter in the NEP between the pixels is small, which is a result of the good fabrication control resulting in a very limited spread of the aluminium properties over the wafer. The amplitude NEP is in most cases a bit higher, and limited for most pixels by the readout noise, which causes the higher value and larger spread. Both NEP values are in excellent agreement with the optical NEP presented in section \ref{section:Sensitivity} which confirms that the dark NEP is a good measurement of the detector sensitivity and that we can expect that a full size lens array coupled to the presented chip would result in a imaging array with a limiting sensitivity given by Fig.  \ref{Fig:DarkNEP}. 

\subsection{Crosstalk}\label{section:Crosstalk}
%==========================================================================================================================
\begin{figure*}
	%\begin{minipage}{1\textwidth}
	\centering
	\includegraphics[width=0.95\textwidth,keepaspectratio]{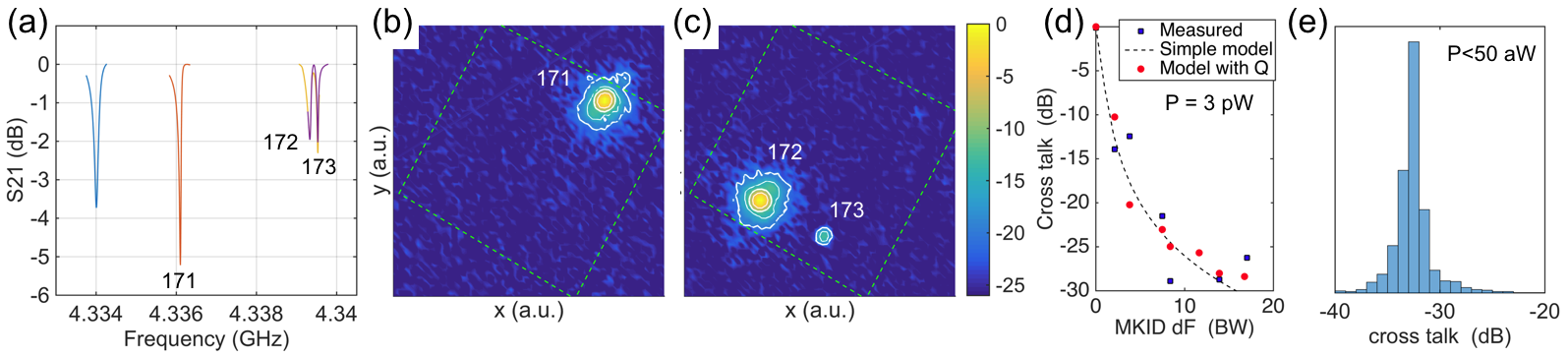}
	%\end{minipage}
	\caption{(a) Frequency sweep ('local sweep') for four resonators, of which the last two are very close together. (b) Phase response of a tone placed at MKID 171 to the position of a point source at the image plane of the chip outside the cryostat; we see a single beam response. (c) Phase response of a tone placed at MKID 172 to the position of a point source at the image plane of the chip outside the cryostat; we see a response of the pixel itself and a -10-dB response at the spatial position of MKID 173. (d) Measured (blue squares) and modelled crosstalk (red dots) together with a simple model calculation (dashed line) as guide to the eye. (e) Estimated crosstalk for the array under dark and cold conditions obtained from a model calculation based upon the data presented in Fig. \ref{Fig:Fsweep}(a).}
\label{Fig:CrossTalk}
\end{figure*}
%=========================================================================================================================
In any imaging system it is important that the spatial information in the scene being viewed is transferred with high fidelity into the final image. This requires that each pixel responds only to its position in the re-imaged focal plane of the system. Unwanted reflections and insufficient baffling in the optical system can cause ghosting and stray light.  In addition, an individual detector pixel can exhibit spatial crosstalk, whereby its signal is partly dependent on the signals of other pixels and vice versa. Both effects limit the image quality and can even make high fidelity observations impossible. Here we discuss crosstalk between detector pixels as this is a problem associated with the detector chip and its readout system rather than with the optical system. For an MKID array we can identify three sources of pixel-pixel crosstalk:
  \begin{itemize}
      \item Electromagnetic (EM) cross-coupling on the detector chip: MKIDs are resonant circuits that can form coupled resonant circuits, resulting in two resonators (or more) responding to the radiation absorbed by only one pixel \citep{Noroozian2012}. EM cross-coupling decreases with increasing pixel-pixel distance and with increasing difference in resonant frequency. \citet{Yates2014} have shown that pixel-pixel crosstalk can be efficiently minimised by placing resonators that are neighbours in frequency space at least one resonator apart. This design criterion is implemented in our array as shown in Fig. \ref{Fig:1}(a).
       \item Crosstalk in the readout due to overlapping resonances: When two resonances are very close in frequency space the change in frequency of one resonator will affect the measured transmission at the resonant frequency of the other resonator, see \citet{Adane2016}. It can be mitigated by designing the resonators so that their resonant frequencies are sufficiently far away in frequency space. In the fabrication process there is some inevitable scatter in the resonant frequencies of the MKIDs due to thickness variations of the superconducting films, especially the NbTiN film in our case. Note that this effect gives crosstalk only between 1-2 pixels, which is naturally suppressed when co-adding the pixel responses to create a final image. 
       \item Surface waves due to re-scattered radiation: In any antenna-coupled imaging system such as the MKID array presented here, not all radiation falling on the detector array will be absorbed. The fractional surface area covered with lenses is not 100\% and not all power falling on each lens is absorbed due to the single mode nature of the radiation coupling. This re-scattered radiation can create  a surface wave inside the chip due to its high dielectric constant. The mesh absorber in our chip is designed to absorb efficiently these surface waves \citep{Yates2017}.
\end{itemize}	
We found that the most effective method to characterise crosstalk is to determine the spatial response of the array pixels using a small calibration source scanning in 2-D in the image plane of the detector array. To do this we use a large cryogenic test facility which can look into the laboratory. It is equipped with an optical system creating a virtually aberration-free image of the detector chip outside the cryostat. The system optics has a Lyot stop at the position of the optics pupil at 4 K resulting in a 1{\it F}$\lambda$ sampling of the detector array, which is a good compromise between resolution and the lowest possible power per pixel. Infrared and FIR filters at various stages define a passband from 825 - 905 GHz, i.e. the same as in the other experiments. This experimental system is described in detail in \citet{Yates2017}. Additionally, we replace the cover of the detector holder with a lid that has a grid of 0.2 mm diameter holes, one in front of each lens. This reduces the power absorbed per pixel from 40 pW to a few pW without reducing the spatial resolution. Note that even in this case the power loading per pixel is many orders of magnitude higher than in the nominal operation condition in space.

Using the same multi-tone readout as for the other experiments we measure the response of all pixels as a function of the position of a 2-mm diameter 900 $^\circ$C calibration source in the re-imaged focal plane of the chip. The results are shown in Fig.  \ref{Fig:CrossTalk}. In panel (a) we see four MKID resonances, of which two are spaced very closely together in frequency and the two others are spaced significantly further apart. In \ref{Fig:CrossTalk}(b) we show the response of MKID 171 to a 2-D scan of the calibration source. MKID 171 is a few bandwidths away from the nearest pixels in frequency space. We observe that MKID 171 responds only to a single spatial position of the source. Additionally, it has a clean beam pattern and we see no evidence of ghosting, crosstalk or stray light issues within the dynamic range of the measurements, which is -30 dB below the pixel peak response. In panel (c) we plot the response of MKID 172. Here we see that this pixel respond to two positions, a main response at the pixel position and a response at a -10-dB level at the spatial location of pixel 173. This is crosstalk due to overlapping resonators. By assessing the patterns for all of the 181 pixels which can be illuminated we see this type of crosstalk only for resonators very close in frequency space, i.e. due to overlapping resonators. There is no evidence of EM cross coupling on this array. This is the same conclusion as in \citet{Yates2014}. 

%==========================================================================================================================
\begin{figure*}
	%\begin{minipage}{1\textwidth}
	\centering
	\includegraphics[width=0.95\textwidth,keepaspectratio]{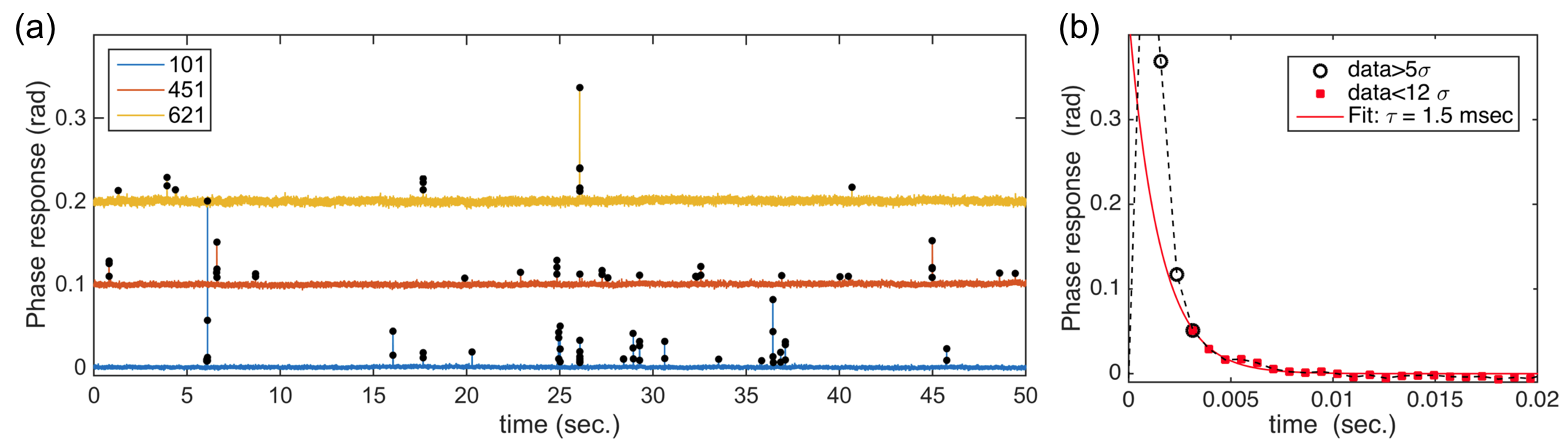}
\caption{(a) Phase response of three MKIDs at 1.2 kHz data rate, measured over a 50-s interval. The black dots represent data with a magnitude $>$ 5 $\sigma_{RMS}$ identified as part of a glitch. (b) A single glitch and an exponential fit to the $<$12$\sigma_{RMS}$ glitch profile, which yields the quasiparticle lifetime.} 
\label{Fig:CR}
\end{figure*}
%==========================================================================================================================
To further investigate the nature of the crosstalk within the array we plot, in Fig. \ref{Fig:CrossTalk}(d) the measured crosstalk between all cross-talking pixel as a function of resonator bandwidth. We over-plot the result with two models that describe the crosstalk between pixels assuming only overlapping resonator crosstalk and also assuming a perfect Lorentzian resonance feature shape \citep{Bisigello2016}. Using a very simple model, based upon the average Q factor of all resonators in the array and the measured resonance frequency difference between the two resonators we observe a reasonable agreement with the measurement. Taking into account the actual Q factors of the 2 cross talking MKIDs under consideration the agreement improves slightly. We can use this model, in combination with the data given in Fig. \ref{Fig:Fsweep}(a) to calculate the expected crosstalk due to overlapping resonators under 'cold and dark' conditions. The result is given in Fig. \ref{Fig:CrossTalk}(e). We observe a crosstalk level well below -30 dB, with 829 of the pixels below this value. The level of crosstalk is much lower than in Fig. \ref{Fig:CrossTalk}(d) because the power per pixel in panel (d) is in excess of several pW, which significantly increasing the MKID bandwidth and hence the cross talk. Under operation in an astronomical observatory as listed in Table \ref{table:1} the absorbed power from the background can only be as high as 130 fW; This power level does not significantly change the resonator bandwidth compared to the cold and dark value; hence the results of Fig. \ref{Fig:CrossTalk}(d) can be taken as nominal for the MKID array in a practical instrument. 

\subsection{Cosmic ray effects} \label{Section:Cosmic rays}

Space observatories operating outside low-Earth orbit are subject to interactions with high energy particles, primarily protons, normally referred to as cosmic rays. These particles are so energetic that it is impossible to construct effective shielding for the detector arrays, and hence the cosmic rays will inevitably interact with the detector chip thereby depositing a fraction of their energy by ionisation and atomic excitation. We have simulated the cosmic ray interaction in our detector chip, taking into account the geometry of the cryostat, and found that the energy deposited has a broad spectrum that peaks at 200 keV.\footnote{For the cosmic ray simulation we used the CRY database http://nuclear.llnl.gov/simulation/ and the GEANT4 simulation toolkit https://geant4.web.cern.ch/geant4/.}. The count rate in an L2 orbit, as measured by \citet{Planck2014}, is 5 events/s/cm$^2$. The typical result of a cosmic ray interaction is a glitch that recovers within a timescale that depends on the exact details of the detector system. Operation of the Planck HFI instrument at L2, has shown that cosmic ray events result in difficult data retrieval and loss of integration efficiency \citep{Planck2014}. In the case of the MKID detector array discussed in this paper it is possible that, in the absence of any countermeasures, a single interaction on the detector chip will result in a glitch visible over the entire area of the detector chip: \citet{Swenson2010} has shown that this is exactly what happens in a small array of MKIDs. \citet{Monfardini2016} recently demonstrated that it is possible to harden MKID arrays against cosmic ray events by adding a layer of a superconducting material with a critical temperature below that of the aluminium of the MKIDs: The general idea is that non-thermal (high-energy) phonons, created by the initial interaction and subsequent phonon down-conversion, are converted to phonons with an energy $E<2\Delta_{Al}$ through electron-phonon interactions in the low-temperature superconducting layer. The MKID array presented here has a low {\it Tc} sub-stoichiometric TiN mesh to absorb stray radiation as discussed in section \ref{Sec:DetDes} and shown in Fig. \ref{Fig:1}(d). This layer doubles as the downconverter for high energy phonons. 

We evaluate the effect of cosmic ray interactions in the detector chip by measuring the effects of secondary cosmic rays. Secondary cosmic rays result from the interaction of primary cosmic rays with the Earth's atmosphere and are easily measured; we have to remove the resulting glitches in all our experiments to obtain the results presented so far. The de-glitching scheme we use consists of three steps: i) We use a simple baseline subtraction to remove all slow drifts in the data by applying a 1-second low pass filter and subtracting the filtered data from the original. This is needed only in cases where we modify the black body or chip temperature over time. ii) We calculate the rms noise for a subset of the corrected data not affected by glitches. iii) We identify each data point with a signal $>$ 5$\sigma_{RMS}$ as affected by a glitch. These data points are removed from the original dataset to create de-glitched data. In the experiment described here we want to study the cosmic ray effects in detail. For this purpose we take 30 minutes of data with the readout system in its fast, low resolution setting with a data rate of 1.2 kHz while operating the chip in dark, cold conditions. A typical result for three MKIDs and 50 seconds of data is shown in Fig. \ref{Fig:CR}(a) and a single pulse is shown in Fig. \ref{Fig:CR}(b).

Referring to Fig. \ref{Fig:CR}(a) we see that the timelines of the MKIDs are affected by the glitches and we observe that there is no 1:1 correlation between the time streams, i.e. not all glitches are seen by all pixels, which is evidence that the mesh absorber is indeed preventing the spread of high-energy phonons over the entire chip. \citet{Catalano2016} reached the same conclusion, and in addition they proved by simulations that the residuals of the cosmic rays at levels $<$ 5$\sigma$ present in the data add only a few \% to the RMS noise and that the residual non-Gaussian features do not affect the integration efficiency. This argument allows us to treat cosmic ray interactions in terms of a loss of integration time, which is taken here as the fraction of data points with a value $> 5\sigma_{RMS}$. We find a fractional loss of integration time of 3.2$\times$10$^{-4}$ and an event rate of 0.18 events/s per MKID. When considering all glitches on the entire chip we find an event rate of 1.3 events/s (425 events/s/m$^2$), which implies that, on average, only $\sim$14\% of the MKIDs are affected by each cosmic ray interaction on the chip. This has the important consequence that the mean cosmic ray dead time per pixel  is \emph{independent} of the chip area. We can estimate the effect of cosmic ray interactions when operating the array in L2 by scaling the hit rate on the chip to the measured event rate from Planck of 5 events/s/cm$^2$. This simple scaling results in an estimated loss in integration time of 4\%. The main uncertainty in this number arises from the spectral energy difference of the power absorbed in the chip between primary and secondary cosmic rays. Dedicated experiments are required to address this issue. From Fig. \ref{Fig:CR}(b) it is clear that we resolve the glitch fully in time with the 1.2-kHz sampling speed. We can even obtain the quasiparticle lifetime from a single exponential fit of the pulse tail, which results in $\tau_{qp}$ = 1.5 msec., similar to the values found by \citet{Visser2011}. The first part of the pulse is associated with a faster decay, which is a result of $\tau_{qp}$ decreasing with increasing quasiparticle density \citep{Baselmans2008}. 

The results so far are obtained under dark, cold conditions representative of an imaging detector array operating in a cryogenically cooled observatory. For the case of a CMB mission the power loading per pixel is significantly higher, on the order of 100 fW per pixel. To be able to evaluate statistically the effect of cosmic ray interactions we increase the bath temperature to 260 mK under dark conditions, thus simulating the effect of a homogeneous illumination of the array which is not possible in our measurement setup. Under these conditions we find that the loss of integration time is reduced to 9$\times$10$^{-5}$, which would result in L2 in a data loss of approximately 1\%. This reduction in cosmic ray susceptibility at a higher power loading of the MKIDs occurs because of the reduction in the quasiparticle lifetime with increasing loading power \citep{Visser2014}.

\subsection{Yield} \label{Section:Yield}

The yield of the system is given by the fraction of pixels that satisfy the key requirements set out in Table \ref{table:2}. In the experiments presented here we have been able to measure, for all pixels individually, the intrinsic sensitivity, the crosstalk, and the cosmic ray dead time. The other parameters have been obtained only for the central pixels. Referring to Table \ref{table:2} and the results presented we evaluate the yield based upon goal parameters except the NEP:
\begin{itemize}
\item $\mathrm{ NEP_{det}<5\times10^{-19}{\WHz}}$ (baseline)
\item Crosstalk $<$ -30 dB (goal)
\item Cosmic ray dead time estimation for operation in L2 $<$ 10\% (goal)
\end{itemize}
We find that 799 out of the 961 pixels meet the above requirements, which implies a yield of 83\% for the detector system. An improvement in yield is possible by using more homogeneous NbTiN films. In the current device the NbTiN increases in resistivity and reduces in thickness from the chip centre to the chip edge, this is similar to the effect described in \citet{Adane2016}. Recent work from \citet{Thoen2016} addresses this problem by using a large-target reactive sputtering system.
%------------------------------------------------------------------------------------------------------------------------------------------------------------------------------------
\begin{table*}
\caption{Measured parameters for the demonstrator system. The 1/{\it f} knee is given for 'cold and dark' conditions, applicable for low-background imaging systems, and for a 56 fW power per pixel, which is the estimated low-power limit for a CMB mission. }              % title of Table
\label{table:3}      % is used to refer this table in the text
\centering                                      % used for centering table
\begin{tabular}{c c c c c c c c c c c}          % centered columns (4 columns)
\hline\hline                        % inserts double horizontal lines \textbf{$\lambda/\Delta\lambda$}
 \textbf{ pixels} & \boldmath$\lambda$  	& \boldmath$\lambda/\Delta\lambda$& \boldmath$\mathrm{NEP_{det}}$	&  \textbf{Absorption}	& \textbf{Instantaneous}	& \textbf{Cosmic Ray}	&  \textbf{Crosstalk}	& \textbf{1/{\it f} knee}& \textbf{1/{\it f} knee}&\textbf{Yield}\\  	
 			& 			 		& 							& 							&  \textbf{efficiency}		& \textbf{dynamic range$^d$}	& \textbf{dead time}		&  				& \textbf{phase} & \textbf{amplitude}& \\  	 
 961			& 350 $\mu$m			&  5							& 3$\times 10^{-19}{\WHz}$		& \textbf{$>$0.72$^{a}$}		& $\mathrm{1\times10^5}$	& $<$5\%	 			&	$<$-30 dB		& n.a.${^b}$  	&$<$0.05		&$>$83\%	\\  	
			& 					&  							& 							&					& 					& 			 		&				& 0.2Hz$^c$	&$<$0.05$^c$	&	\\ 							
\hline                                             %inserts single line
\hline  
\end{tabular}
\tablefoot{
$^{a}$ The absorption efficiency is referred to an optimal design with a single polarisation input which has an aperture efficiency $\mathrm{\eta_{AE} = 0.8}$. The measured aperture efficiency $\mathrm{\eta_{AE} = 0.58}$.
\\ 
$^{b}$ The spectra at 'dark and cold' conditions (10 aW/pixel) are not white.
\\
$^{c}$ Measured at 50 fW power per pixel.
\\
$^d$ The instantaneous dynamic range is defined as the ratio of the maximum source power divided by the limiting NEP under dark and cold conditions.
}
\end{table*}

%-------------------------------------------------------------------------------------------------------------------------------------------------------------------------------------
\section{Discussion}%%%%
\label{Sec:Dis}
The results presented in this paper show that we have achieved our goal to create an imaging system that combines the high sensitivity obtained for a single MKID pixel with a large array that is read out with a single readout system. We summarise the measured results in detail in Table \ref{table:3}; these can be compared to the requirements and goals set in Table \ref{table:2}. We reach the baseline requirements for the sensitivity, wavelength range and bandwidth, and we also reach the goal requirements for the pixel yield, crosstalk, cosmic ray dead time and absorption efficiency. Only for the 1/{\it f} knee frequency are the requirements not yet met for the lowest backgrounds. When comparing these results to the real mission requirements given in Table \ref{table:1} we see that the a system such as the one presented here could be used for a CMB instrument or for a camera instrument on a 25 K telescope. The only substantial development required is to demonstrate scaling to longer wavelengths for the CMB instrument, and at shorter wavelengths for the camera instrument. For a camera on a 5K telescope both a small improvement in sensitivity and improved 1/{\it f} performance are needed additionally.

\subsection{Further improvements of the detector array}

The low-background sensitivity of the detectors at $\mathrm{P_{opt}<}$ 0.1 fW is limited by TLS noise using phase readout and readout noise when using amplitude readout, i.e. we are detector-limited. Possible solutions are i) decreasing the device TLS noise by reducing the noise or increasing the responsivity, and ii) increasing the device {\it Q} factor, which results in a higher MKID output amplitude noise, in combination with a readout scheme that uses amplitude readout for small signals and phase readout for larger signals. These could reduce the NEP into the range of few 10$^{-19} \WHz$ down to 0.1 Hz. Reaching the sensitivity requirements for a spectroscopic instrument with a cooled telescope (requiring NEP$=1\times10^{-20}\;\WHz$) would demand a much more dramatic optimisation. 

Increasing the operating FIR frequency and bandwidth is possible but requires a modification of the MKIDs: the design and material choice of the device presented in this paper limit the operation frequency to values below 1.1 THz. Higher operational frequencies are possible by using aluminium for the antenna structure and the ground plane of the narrow section of the MKID. In \citet{Visser2014} a thick ground plane and a thinner central line of aluminium were used to minimise the loss of signal in the ground plane to 10\%. Additionally, a broader detection bandwidth is possible with a similar detector geometry by using a different antenna. Recently \citet{Bueno2017} have shown similar sensitivities to those presented here at 1.4-2.8 THz using a leaky lens-antenna coupled MKID \citep{Neto2014}. 

\subsection{Towards space qualification of MKID arrays}
For implementation in a future satellite mission the detector arrays will have to be qualified for space. The  arrays are relatively simple and robust structures made out of mechanically stable, solid substrates and from relatively thick metallic films which should not pose any fundamental issues regarding vibration, acoustic and thermal requirements. Likewise, although full evaluation needs to be carried out, there is good reason to believe that there will be no particular issues with degradation in performance associated with long-term accumulated ionising radiation dosage or with ability to survive multiple thermal cycles, storage for long periods. The effects of ionising radiation on data quality have been studied by \citet{Karatsu2016} for all aluminium devices with an identical quasiparticle lifetime and found to be negligible. Additionally the effects of cosmic rays on large MKID arrays are still relatively poorly studied. Further experiments are needed to provide conclusive evidence that the very positive results presented here are fully representative for real operation in space.

\subsection{Towards space qualification of the MKID readout system}

The readout system presented here \citep{Rantwijk2016} is designed for flexibility and made from commercial components for laboratory use. Radiation-hard versions exist for all critical components of the readout system, making it possible to design a space qualified version. For example, we can  design a digital back-end based on: i) the E2V EV12DS130A as DAC, ii) TI ADC10D1000QML-SP as ADC, iii) the XILINX Virtex-5 XQR5VFX130 FPGA for the data processing which requires additionally iv) QDR memory chips: Cypress CYRS1544AV18 QDR. The estimated power consumption for such a readout system is of the order of 15 - 25 mW/pixel, dissipated at ambient temperature, assuming a 70\% point of load efficiency and depending strongly on the exact requirements for the final instrument. A single cryogenic HEMT LNA similar to the one used in this paper \citep{Diez2003} must be used. The nominal power dissipation is 5 mW at 4 K, operation at higher temperatures could be considered with negligible, or limited loss in system sensitivity.

\section{Conclusions}
\label{Sec:Con}
We have demonstrated an MKID-based imaging system consisting of a 961 pixel MKID array that is read out using a single readout system and one pair of readout cables. The readout requires one low noise amplifier operating at $\sim$4K with a power dissipation of a few mW. This demonstrator represent a major step forward in FIR detector technology, especially in terms of multiplexing capabilities in combination with a very high sensitivity. It fulfils many generic requirements for future space based observatories. The detector array operates in a 20\% bandwidth around 850 GHz. The sensitivity obtained with a thermal calibration source is given by $\mathrm{<NEP_{det}>=3\times10^{-19}\WHz}$ with an aperture efficiency of 0.58, which represents 73\% of he theoretical limit for a single mode system. Furthermore we achieve 83\% yield, low crosstalk ($<$-30dB) and a dynamic range of $10^5$, enabling measurement of sources of close to 1 Jy for most applications. Additionally the detector array is hardened against cosmic ray interaction with an expected loss of integration time of less that 4\% when operated in L2. This proves that MKID technology is now sufficiently mature for consideration in future space based observatories.

\acknowledgements{We wish to thank L. Bisigello and D. Flanigan for useful discussions. This work was supported as part of a collaborative project, SPACEKIDs, funded via grant 313320 provided by the European Commission under Theme SPA.2012.2.2-01 of Framework Programme 7. The contribution of J.J.A. Baselmans is also supported by the ERC COG 648135 MOSAIC. J. Martin-Pintado acknowledges partial support by the MINECO under grants AYA2010- 2169-C04-01, FIS2012-39162-C06-01, ESP2013-47809-C03-01 and ESP2015-65597, A.B. Baryshev is supported also by ERC STG 240602 TFPA 
}

\bibliographystyle{aa}
\bibliography{Jochem_biblio}

\end{document}